\documentclass[11pt]{article}
\usepackage{lmodern}
\usepackage{lmodern}
\newcommand{\comment}[1]{ }
\usepackage[utf8]{inputenc}
\usepackage[a4paper]{geometry}
\usepackage{color}
\usepackage{float}
\usepackage{amsmath}
\usepackage{amssymb}
\usepackage{graphicx}
\usepackage[numbers,square,numbers,comma,sort&compress]{natbib}
\usepackage[unicode=true,
 bookmarks=false,
 breaklinks=false,pdfborder={0 0 0},pdfborderstyle={},backref=false,colorlinks=true]
 {hyperref}
\hypersetup{
 citecolor=blue,linkcolor=blue,urlcolor=blue}

\makeatletter

\providecommand{\tabularnewline}{\\}

\usepackage[square,numbers,comma,sort&compress]{natbib}
\usepackage{slashed}
\def\varabstract{ }
\def\varkeywords{ }
\def\vararxivnumber{ }
\def\vartitle{ }
\def\varpreprint{ }
\renewcommand{\title}[1]{\gdef\vartitle{#1}}
\renewcommand{\abstract}[1]{\gdef\varabstract{#1}}
\newcommand{\keywords}[1]{\gdef\varkeywords{#1}}
\newcommand{\arxivnumber}[1]{\gdef\vararxivnumber{#1}}
\newcommand{\preprint}[1]{\gdef\varpreprint{#1}}
\newtoks\authtoks
\renewcommand{\author}[2][]{%
	\authtoks=\expandafter{\the\authtoks#2$^{#1}$\ }%
}
\newtoks\affiltoks
\newcommand{\affiliation}[2][]{%
    \affiltoks=\expandafter{\the\affiltoks{\item[$^{#1}$]#2}}%
}
\newtoks\emailtoks\newcounter{emailcounter}%
\newcommand{\emailAdd}[1]{%
\ifnum\theemailcounter>0\emailtoks=\expandafter{\the\emailtoks, \typeemail{#1}}%
\else\emailtoks=\expandafter{\typeemail{#1}}%
\fi
\stepcounter{emailcounter}}
\newcommand{\typeemail}[1]{\href{mailto:#1}{\tt #1}}

\renewcommand{\maketitle}{
	\newgeometry{margin=2cm}
	\pagestyle{empty}\setcounter{page}{0}
	\if!\varpreprint!\else\begin{flushright}\varpreprint\end{flushright}\fi
	{\huge\flushleft\sffamily\bfseries\vartitle\par}
\vskip6ex
{\large\bfseries\raggedright\sffamily\the\authtoks\par}
\vskip2ex
\begin{list}{}{%
\setlength{\leftmargin}{0.28cm}%
\setlength{\labelsep}{0pt}%
\setlength{\itemsep}{-3pt}%
\setlength{\topsep}{-\parskip}}
\itshape\small%
\the\affiltoks
\end{list}
\vskip2ex
\noindent\hspace{0.28cm}\begin{minipage}[l]{.9\textwidth}
\begin{flushleft}
\textit{E-mail:} \the\emailtoks
\end{flushleft}
\end{minipage}
\vskip5ex
\noindent{\renewcommand\baselinestretch{.9}\textsc{Abstract:}}\ \varabstract
\vskip5ex 
\if!\varkeywords!\else\noindent{\textsc{Keywords:}}\ \varkeywords \vskip2ex\fi
\if!\vararxivnumber!\else\noindent{\textsc{ArXiv ePrint:}} \href{http://arxiv.org/abs/\vararxivnumber}{\vararxivnumber}\vskip2ex\fi
\newpage
\restoregeometry
\pagestyle{plain}
\hrule
\bigskip\bigskip

{
	\hypersetup{linkcolor=black}
	\tableofcontents
}
\bigskip\medskip
\hrule
\bigskip\bigskip
\setcounter{footnote}{0}
}

\title{Dark matter from a complex scalar singlet:\\ The role of dark CP and other discrete symmetries}

\author[a]{Leonardo Coito,}
\author[a]{Carlos Faubel,}

\author[a]{Juan Herrero-Garc\'ia,}
\author[a]{Arcadi Santamaria}
\affiliation[a]{Departament de F\'isica Te\`orica, Universitat de Val\`encia and IFIC, Universitat de Val\`encia-CSIC\\ Dr. Moliner 50, E-46100 Burjassot (València), Spain}

\emailAdd{leonardo.coito@uv.es}
\emailAdd{carlos.faubel@uv.es}
\emailAdd{juan.herrero@ific.uv.es}
\emailAdd{arcadi.santamaria@uv.es}

\abstract{
We study the case of a pseudo-scalar dark matter candidate which emerges from a complex scalar singlet, 
charged under a global U(1) symmetry, which is broken both explicitly and spontaneously. The pseudo-scalar is naturally 
stabilized by the presence of a remnant discrete symmetry: dark CP. We study and compare the phenomenology of several 
simplified models with only one explicit symmetry breaking term.
We find that several regions of the parameter space are able to reproduce the observed dark matter abundance 
while respecting direct detection and invisible Higgs decay limits: in the resonances of the two scalars, 
featuring the known as forbidden or secluded dark matter, and through non-resonant Higgs-mediated annihilations. 
In some cases, combining different measurements would allow one to distinguish the breaking pattern of the symmetry. Moreover, this setup admits a light DM candidate at the sub-GeV scale.  
We also discuss the situation where more than one symmetry breaking term is present. In that case, 
the dark CP symmetry may be spontaneously broken, thus spoiling the stability of the dark matter candidate. 
Requiring that this does not happen imposes a constraint on the allowed parameter space. Finally, we consider an effective field 
theory approach valid in the pseudo-Nambu-Goldstone boson limit and when the U(1) breaking scale is much larger than 
the electroweak scale.
}

\arxivnumber{2106.05289}
\preprint{FTUV-21-0608.9540, IFIC/21-20}

\@ifundefined{showcaptionsetup}{}{%
 \PassOptionsToPackage{caption=false}{subfig}}
\usepackage{subfig}
\makeatother

\begin{document}
\global\long\def\vs{v_{s}}%
\global\long\def\mdm{m_{\theta}}%
\global\long\def\mdmsq{m_{\mathrm{\theta}}^{2}}%
\global\long\def\gev{\mathrm{GeV}}%
\global\long\def\mev{\mathrm{MeV}}%
\global\long\def\tev{\mathrm{TeV}}%

\global\long\def\sa{s_{\alpha}}%
\global\long\def\ca{c_{\alpha}}%

\global\long\def\Real#1{\mathrm{Re}\left\{  #1 \right\}  }%
\global\long\def\Imag#1{\mathrm{Im}\left\{  #1 \right\}  }%
\global\long\def\abs#1{\left|#1\right|}%

\maketitle

\section{Introduction \label{sec:intro}}

In the last hundred years very strong evidence for the presence of
non-baryonic cold dark matter (DM) in the Universe has been established
\citep{Jarosik:2010iu,Clowe:2006eq} (see also Refs.~\citep{Bertone:2016nfn,Bertone:2010zza,Bertone:2004pz,Olive:2003iq}
for reviews). However, the Standard Model (SM) does not provide a
suitable particle that could account for it. One of the simplest extensions
of the SM that contains a DM candidate is obtained by adding a real
scalar singlet, $\phi$, equipped with a discrete symmetry that prevents
its decay \citep{Silveira:1985rk,McDonald:1993ex}\footnote{Actually in Ref.~\citep{McDonald:1993ex} the case of $N$ complex
singlets was studied.}. The scalar potential of the SM is
\begin{equation}
V_{{\rm SM}}=m_{H}^{2}H^{\dagger}H+\lambda_{H}(H^{\dagger}H)^{2}\,,\label{eq:sm-potential}
\end{equation}
 where $H$ is the usual Higgs doublet, which after electroweak spontaneous
symmetry breaking can be written in the unitary gauge as
\begin{equation}
H=\frac{1}{\sqrt{2}}\begin{pmatrix}0\\
v+h^{\prime}
\end{pmatrix}\,.\label{eq:Hunit}
\end{equation}
Here $v=246\,\gev$ is the electroweak vacuum expectation value and
$h'$ is the Higgs boson (in the absence of mixing with other scalars).
This potential is now extended by adding three new couplings,
\begin{equation}
\Delta V(H,\phi)=m^{2}\phi^{2}+\lambda_{H\phi}|H|^{2}\phi^{2}+\lambda_{\phi}\phi^{4}\,,\label{eq:one-scalar-potential}
\end{equation}
where terms odd in $\phi$ are removed by imposing the discrete symmetry
$\phi\rightarrow-\phi$, which makes the $\phi$ particles stable
if the symmetry is not broken spontaneously.

The Higgs portal term, $\lambda_{H\phi}|H|^{2}\phi^{2}$, controls
the annihilation of pairs of $\phi$ particles into SM states. The
observed relic abundance can be reproduced via freeze-out for particular
values of the DM mass (in the Higgs boson resonance, or for large
values $m_{\phi}\gtrsim500$~GeV \citep{Burgess:2000yq,Beniwal:2020hjc,Athron:2018hpc,Cline:2013gha}).
The simplest model has been extensively studied \citep{Wu:2016mbe,Han:2015dua,Feng:2014vea,Buckley:2014fba,Cline:2013gha,Guo:2010hq,He:2009yd,Barger:2007im,Burgess:2000yq},
and has been embedded in larger schemes that solve other problems
of the SM such as neutrino masses, inflation and baryogenesis \citep{vanderBij:2006ne,Davoudiasl:2004be}.
The mechanism is quite predictive because exactly the same coupling
with the Higgs field that allows for DM annihilations gives rise to
DM scatterings in direct detection (DD) experiments \citep{Goodman:1984dc}
(see also \citep{Schumann:2019eaa,Liu:2017drf,Undagoitia:2015gya,Baudis:2014naa})
and, if the DM is light enough, induces invisible Higgs decays into
DM \citep{Pospelov:2011yp,Cai:2011kb,Bento:2001yk}. Neither of the
two effects has been observed and this sets important constraints
on the model; in particular, DD experiments have left little parameter
space for this scenario (see for instance Refs.~\citep{Beniwal:2020hjc,Athron:2018hpc,Athron:2018ipf,Escudero:2016gzx}). 

The simplest extension of the model consists in just adding a second
scalar singlet with the appropriate symmetries \citep{Casas:2017jjg,Bauer:2016gys,Buckley:2014fba,Bhattacharya:2016ysw,Drozd:2011aa,Cao:2007fy}.
The most general terms one can add to the SM potential that describe
the interactions of the two scalar singlets can be written as
\begin{equation}
V(H,\phi_{1},\phi_{2})=\frac{|H|^{2}}{\sqrt{2}}(\omega_{1}\phi_{1}+\omega_{2}\phi_{2})+\frac{|H|^{2}}{2}(\alpha_{1}\phi_{1}^{2}+\alpha_{12}\phi_{1}\phi_{2}+\alpha_{2}\phi_{2}^{2})+V_{{\rm I}}+V_{{\rm II}}+V_{{\rm III}}+V_{{\rm IV}}\,,\label{eq:2scalar-potential}
\end{equation}
with
\begin{align}
V_{{\rm I}} & =\frac{1}{\sqrt{2}}\left(\delta_{1}\phi_{1}+\text{\ensuremath{\delta_{2}\phi_{2}}}\right)\,,\label{eq:2scalar-potential-V1}\\
V_{{\rm II}} & =\frac{1}{2}\left(m_{1}^{2}\phi_{1}^{2}+2m_{12}^{2}\phi_{1}\phi_{2}+m_{2}^{2}\phi_{2}^{2}\right)\,,\label{eq:2scalar-potential-V2}\\
V_{{\rm III}} & =\frac{1}{2\sqrt{2}}\left(\mu_{1}\phi_{1}^{3}+\mu_{12}\phi_{1}^{2}\phi_{2}+\text{\ensuremath{\mu_{21}}\ensuremath{\phi_{2}^{2}\text{\ensuremath{\phi_{1}}}+}\ensuremath{\mu_{2}\phi_{2}^{3}}}\right)\,,\label{eq:2scalar-potential-V3}\\
V_{{\rm IV}} & =\frac{1}{4}\left(\lambda_{1}\phi_{1}^{4}+\beta_{12}\phi_{1}^{3}\phi_{2}+\text{\ensuremath{\lambda_{12}}\ensuremath{\phi_{1}^{2}\text{\ensuremath{\phi_{2}^{2}}}+}\ensuremath{\beta_{21}\phi_{2}^{3}\phi_{1}}+\ensuremath{\lambda_{2}\phi_{2}^{4}}}\right)\,.\label{eq:2scalar-potential-V4}
\end{align}

The potential above, which contains 19 real parameters, does not have
any additional symmetry beyond the SM ones. However, the kinetic terms
of the two new scalars do have an $O(2)$ rotation symmetry and a
shift symmetry $\phi_{1,2}\rightarrow\phi_{1,2}+c_{1,2}$. These symmetries
can be used to remove the quadratic mixing $m_{12}^{2}\phi_{1}\phi_{2}$
in $V_{{\rm II}}$ and the linear terms in $V_{{\rm I}}$ (or the
terms $|H|^{2}(\omega_{1}\phi_{1}+\omega_{2}\phi_{2})$). Therefore,
the general potential adds 16 real physical parameters to the 2 parameters
of the SM potential. Notice that these simplifications can always
be used when the potential does not have any symmetry. However, if
one imposes some symmetry that relates the parameters of the potential
one should start from the most general potential compatible with the
symmetry, check if there are parameters that can be removed by using
the symmetries of the kinetic terms and then check whether the symmetries
are broken or not by the global minimum of the potential. For instance,
if we impose $\phi_{1}\leftrightarrow\phi_{2}$, we have that $m_{1}^{2}=m_{2}^{2}$,
but $m_{12}^{2}\not=0$ is allowed; one could go to a different basis
in which $m_{12}^{2}=0$ but with $m_{1}^{2}\not=m_{2}^{2}$, and
in this basis the symmetry $\phi_{1}\leftrightarrow\phi_{2}$ is hidden.
Similarly, in the symmetry basis linear terms would be allowed, as
long as $\delta_{1}=\delta_{2}$, and then one could ask about the
conditions for this symmetry to be preserved by the vacuum.

If the potential does not have any symmetry the new particles do not
carry any conserved quantum number that prevent them from decaying
and, in general, they cannot be DM. The simplest way to have a sufficiently
long-lived particle and therefore a good DM candidate is that it has
at least one preserved symmetry. There are many symmetries one can
impose; for instance, one can require that the potential is invariant
under $\phi_{1}\rightarrow-\phi_{1}$ and $\phi_{2}\rightarrow-\phi_{2}$.
This will remove completely $V_{{\rm III}}$ and the linear terms
in the singlets. Then, if the symmetry is not spontaneously broken,
the lightest of the two new particles will be completely stable. This
model has been recently studied in detail in Ref.~\citep{Casas:2017jjg},
where it has been shown that some of the problems of the model with
just one scalar can be alleviated. One can impose many other symmetries:
$\phi_{1}\rightarrow\phi_{1}$, $\phi_{2}\rightarrow-\phi_{2}$ or
$\phi_{1}\leftrightarrow\phi_{2}$ or complete invariance under the
symmetry of the kinetic term, $O(2)$. Many of these symmetries lead
to equivalent Lagrangians and some may or may not be broken spontaneously.
It is then interesting to classify the different inequivalent possibilities
that could lead to a good DM candidate and study the allowed parameter
space for each of them. Since $O(2)$ is isomorphic to $U(1)$ times
a reflection, it is more natural to explore the symmetries using the
complex parametrization for the two real fields $S\equiv\left(\phi_{1}+i\phi_{2}\right)/\sqrt{2}$.
The possible models that could lead to a stable particle are quite
obvious in this parametrization and can be classified in two groups
depending on whether the complex field $S$ acquires a vacuum expectation
value (VEV) or not. In this work, we focus on the former case because
it helps to avoid DD constraints~\citep{Alanne:2020jwx,Arina:2020alg,Muhlleitner:2020wwk,Gross:2017dan,Coimbra:2013qq}.
Moreover it can also help to alleviate the instability problems of
the scalar potential of the SM (see for instance Ref.~\citep{EliasMiro:2012ay}).
 
The paper is structured as follows. In Sec.~\ref{sec:complex-scalar}
we discuss the general Lagrangian with two real scalar fields by using
the complex parametrization, and we classify all the symmetries one
can impose on it that could lead to a DM candidate. In Sec.~\ref{sec:general-case}
we study the scalar spectrum and couplings in the most general scenario.
We outline the possible explicit symmetry breaking terms, which give
rise to what we term as \textit{minimal models}. In Sec.~\ref{sec:comparison-of-models}
we analyze the DM phenomenology, discussing the relic abundance, DD
and Higgs invisible decays, presenting numerical results for a prototypical
\textit{minimal model} (the quadratic model) and comparing with the
other ones. In addition, we discuss the possibility of a light DM candidate at the sub-GeV scale. In Sec.~\ref{sec:beyond-minimal} we study the case with
several symmetry breaking terms, which in general may lead to the
spontaneous breaking of the symmetry that stabilizes the DM in large
regions of the parameter space. In Sec.~\ref{sec:pGB-limit} we discuss
an effective Lagrangian approach which is appropriate when the DM
candidate is a pseudo-Nambu-Goldstone boson (PNGB). Finally in Sec.~\ref{sec:conc}
we outline the main results of this work. We also provide several
appendices with more technical details.

\section{The complex scalar singlet Lagrangian and its symmetries \label{sec:complex-scalar}}

We write the complex scalar singlet field $S$ as
\begin{equation}
S\equiv\frac{1}{\sqrt{2}}\left(\phi{}_{1}+i\phi_{2}\right)\,.\label{eq:complex-scalar}
\end{equation}
Then, the most general Lagrangian of the SM extended by $S$ can be
written as
\begin{equation}
\mathcal{L}=\mathcal{L}_{{\rm SM}}+|\partial_{\mu}S|^{2}-V(H,S)\,,\label{eq:full-lagrangian}
\end{equation}
where $\mathcal{L}_{{\rm SM}}$ includes the SM potential, $V_{{\rm SM}}$,
in Eq.~\eqref{eq:sm-potential}, and $V(H,S)$ contains all the interactions
of the new complex scalar $S$. In the complex parametrization, the
symmetry of the kinetic term is $S\rightarrow e^{i\alpha}S$ and $S\rightarrow S^{*}$,
namely, $U(1)$ times a reflection, which can be identified with a
``dark'' charge conjugation, or ``dark'' CP (DCP) since the kinetic
term of the scalar also preserves parity.

If the $U(1)$ global symmetry is preserved by the Lagrangian but
is spontaneously broken by the vacuum, there will be an exactly massless
Goldstone boson and, of course, no DM candidate (it would contribute,
though, to dark radiation). Therefore, if there is spontaneous symmetry breaking (SSB), in order to have a DM candidate we need both DCP conservation
and, at least, one term that breaks explicitly the $U(1)$. In this
work, we are interested in this scenario. If the explicit symmetry
breaking is much smaller than the VEV of $S$, the lightest of the
scalars is a PNGB and can be the DM candidate. This limit is studied in Sec.~\ref{sec:pGB-limit}.

In order to have a DM candidate the potential $V(H,S)$ has to preserve,
at least, a discrete symmetry that stabilizes the DM candidate. The
possible discrete symmetries one can impose are the discrete subgroups
of the kinetic term symmetry group, namely, $O(2)\sim U(1)+\mathrm{reflection}$,
compatible with a polynomial Lagrangian. These are $Z_{2}$ ($S\rightarrow-S$),
$Z_{3}$ ($S\rightarrow e^{i2\pi/3}S$) , $Z_{4}$ ($S\rightarrow iS)$
and the reflection, DCP, ($S\rightarrow S^{*}$).\footnote{If $S$ does not get a VEV, all the symmetries discussed can be used
to stabilize the DM. For instance, the case in which only $Z_{2}$
is imposed, $S\rightarrow-S$ and general complex couplings are allowed
has been discussed in Ref.~\citep{Casas:2017jjg}. In Ref.~\cite{Ghosh:2021qbo} the authors have explored the possibility of having asymmetric DM in an extended scenario with a real scalar singlet in addition to the complex scalar singlet.} One may think that there could also be other discrete symmetries
like $S\rightarrow-S^{*}$ ($\phi_{1}\rightarrow-\phi_{1},\phi_{2}\rightarrow\phi_{2}$
in the real parametrization) or $S\rightarrow iS^{*}$ ($\phi_{1}\leftrightarrow\phi_{2}$
in the real parametrization), but all the symmetries that involve
a reflection are physically equivalent to DCP, as can easily be seen
by rephasing the field and redefining the parameters in the Lagrangian.
Therefore, we define DCP as $S\rightarrow S^{*}$. A necessary condition
for DCP conservation is that all couplings in the potential are real,
like in the Standard Model CP. Notice, however, that since the potential
only depends on $|H|^{2}$ this DCP can be defined independently of
the standard CP. Then, it will not be affected by the SM violations
of CP and will not induce additional CP violation in the SM.

Given the possible symmetries, it is natural to split the potential
in the terms that preserve the $U(1)$ symmetry, $V_{0}$, and the
terms that break it explicitly, $V_{1,\mathrm{full}}$, $V_{Z_{2},\mathrm{full}}$,
$V_{Z_{3}}$ and $V_{Z_{4}}$, classified according to the discrete
symmetries they preserve. That is
\begin{equation}
V(H,S)=V_{0}+V_{1,\mathrm{full}}+V_{Z_{2},\mathrm{full}}+V_{Z_{3}}+V_{Z_{4}}\,,\label{eq:complex-scalar-potential}
\end{equation}
with
\begin{equation}
V_{0}=m_{S}^{2}S^{\dagger}S+\lambda_{S}(S^{\dagger}S)^{2}+\lambda_{HS}(H^{\dagger}H)(S^{\dagger}S)\,,\label{eq:symmetric-potential}
\end{equation}
and
\begin{align}
V_{1,\mathrm{full}} & =\frac{1}{2}\mu^{3}S+\frac{1}{2}\mu_{H1}|H|^{2}S+\frac{1}{2}\mu_{1}(S^{\dagger}S)S+\mathrm{H.c.}\,,\label{eq:complex-scalar-potential-V1}\\
V_{Z_{2},\mathrm{full}} & =\frac{1}{2}\mu_{S}^{2}S^{2}+\frac{1}{2}\lambda_{H2}|H|^{2}S^{2}+\frac{1}{2}\lambda_{2}(S^{\dagger}S)S^{2}+\mathrm{H.c.}\,,\label{eq:complex-scalar-potential-V2}\\
V_{Z_{3}} & =\frac{1}{2}\mu_{3}S^{3}+\mathrm{H.c.}\,,\label{eq:complex-scalar-potential-V3}\\
V_{Z_{4}} & =\frac{1}{2}\lambda_{4}S^{4}+\mathrm{H.c.}\,.\label{eq:complex-scalar-potential-V4}
\end{align}
The second, third and fourth terms respect a $Z_{2},Z_{3}$ and $Z_{4}$
symmetry, respectively. The parameters $m_{S}^{2},\text{\ensuremath{\lambda_{HS}}},\lambda_{S}$
in $V_{0}$ are all real, therefore, $V_{0}$ is also invariant under
$S\rightarrow S^{*}$. All the parameters of the symmetry breaking
terms are, in principle, complex. Therefore, the general complex potential
contains 3 real and 8 complex couplings, 19 real parameters, as in
the real parametrization. The correspondence between the couplings
in the real and complex parametrizations is given in Appendix~\ref{sec:Correspondence-re-complex}.
If all the couplings in the complex parametrization are real, the
8 couplings $m_{12}^{2}$, $\delta_{2}$, $\omega_{2}$, $\alpha_{12}$,
$\mu_{2}$, $\mu_{12}$, $\beta_{12}$, $\beta_{21}$ in the real
parametrization vanish. Therefore, the presence of any of the latter
implies that the DCP is explicitly violated in the potential.

By using a phase redefinition of $S$ one can choose one of the couplings,
for instance $\mu_{S}^{2}$, real (and positive or negative), which
would be equivalent to choosing $m_{12}^{2}=0$ in the real parametrization.
Moreover, as in the real parametrization, the linear terms can be
expressed in terms of other couplings when selecting the global minimum
of the theory. Then, the general potential in the complex parametrization
contains 4 real and 6 complex couplings, 16 real physical parameters
in total, like in the real representation. From the 16 parameters,
3 preserve the $U(1)$ symmetry and 13 break it. However, we should
remark that if we are going to impose some symmetry, to count the
number of physical parameters we should start with the general Lagrangian,
including all the terms, and see how many of them survive the symmetry.

If the minimum of the potential is found for $\left\langle S\right\rangle \not=0$,
$U(1)$ undergoes SSB. Then, also $Z_{2}$, $Z_{3}$ and $Z_{4}$
will be broken by the VEV of $S$. However, if all couplings are real,
DCP will only be broken if the VEV of $S$ is complex.

It is easy to see that if there is only one $U(1)$ symmetry breaking
term its coupling can be taken real. Moreover, in that case, the VEV
of $S$ can always be chosen real \citep{Haber:2012np}, therefore
DCP will be automatically preserved also by the vacuum. Alternatively,
if there are several symmetry breaking terms, depending on the values
of the couplings, the VEV can be complex even if the couplings are
real and, therefore, the system can suffer from spontaneous DCP violation
(see Sec.~\ref{subsec:About-Spontaneous-CP}). Then, we arrive at
the conclusion that if there is SSB of $U(1)$ the only symmetry that
could provide a DM candidate is DCP. In this case, the other discrete
symmetries $Z_{2}$, $Z_{3}$, $Z_{4}$ can only be used to simplify
the Lagrangian\footnote{Notice that the spontaneous breaking
of discrete symmetries can lead to domain wall problems~\citep{Zeldovich:1974uw,Vilenkin:1984ib}. However, since in our case these $Z_n$ are only used to simplify the potential, they can be considered as approximate symmetries,
thus avoiding domain wall problems.} and/or to motivate the presence of the DCP symmetry
(e.g. in the case of $Z_{3}$ or $Z_{4}$). 

The general case of SSB of $U(1)$ with all possible explicit symmetry
breaking terms contains many parameters and it is not very predictive.
Moreover, the discrete symmetry DCP, necessary to have a DM candidate,
must be imposed by hand, and even in this case it could be broken
by the vacuum, leading to an unstable pseudo-scalar. On the other
hand, if there is only one explicit $U(1)$ symmetry breaking term,
as discussed above, DCP is automatically conserved. Therefore, it
makes sense to study first the cases with only one symmetry breaking
term and then analyze how the addition of other terms modify the simplest
scenarios. This can also be justified by using symmetries or by taking
the softest symmetry breaking terms. We will make this analysis in
Sec.~\ref{subsec:minimal-models}.

\section{Mass spectrum and couplings \label{sec:general-case}}

\subsection{The general case}

Following the discussion above we will consider the general potential
in Eqs.~(\ref{eq:complex-scalar-potential}--\ref{eq:complex-scalar-potential-V4})
and require a DCP symmetry, \emph{e}.g. that all the couplings in
the potential are real. Then we will have 11 real parameters, 3 preserving
$U(1)$ and 8 breaking it. Notice that, when $S$ is written in terms
of real and imaginary parts, $\phi_{1,2}$, the Lagrangian is invariant
under $\phi_{2}\rightarrow-\phi_{2}$ which is just the manifestation
of the DCP symmetry. If unbroken by the vacuum, this symmetry will
avoid $\phi_{2}$ from decaying and it will be a good DM candidate.
The real part $\phi_{1}$, however, has completely general couplings;
in particular it has linear and cubic terms. Since the potential in
terms of $\phi_{1}$ is the most general potential we can write, it
has the same form after a shift transformation $\phi_{1}\rightarrow\phi_{1}+\sigma$
which can be used to remove one of the parameters, for instance the
cubic term $\mu_{H1}|H|^{2}S$ or the linear term $\mu^{3}S$, leaving
only $10$ real parameters. However, for the time being, we maintain all of them, keeping in mind that if all of the DCP conserving couplings
are present one of them is redundant. Then, we will assume that the
scalar singlet also takes a VEV, $v_{s}$. Using the linear parametrization
for the singlet, we have
\begin{equation}
S=\frac{1}{\sqrt{2}}\left(\vs+\text{\ensuremath{\rho^{\prime}}}+i\theta\right)\,.\label{eq:fields-with-vevs}
\end{equation}
The Higgs doublet VEV, $v$, can always be taken real, and to preserve
the DCP symmetry we will require that also $\vs$ is real. Notice
that the most general potential could lead to a complex $v_{s}$,
therefore breaking spontaneously the DCP and spoiling the stability
of the DM candidate. In Sec.~\ref{subsec:About-Spontaneous-CP} we
will comment on the conditions to avoid spontaneous DCP violation.

Using the minimization equations for $h^{\prime}$ and $\rho^{\prime}$,
the bare mass parameters can be written in terms of the couplings
and VEVs,
\begin{align}
-m_{H}^{2} & =\frac{1}{2}(\lambda_{H2}+\lambda_{HS})v_{s}^{2}+\lambda_{H}v^{2}+\frac{\sqrt{2}}{2}\mu_{H1}v_{s}\,,\\
-m_{S}^{2} & =\mu_{S}^{2}+(\lambda_{2}+\lambda_{S}+\lambda_{4})v_{s}^{2}+\frac{1}{2}(\lambda_{H2}+\lambda_{HS})v^{2}+\frac{\sqrt{2}}{4}\mu_{H1}\left(\frac{v}{v_{s}}\right)v\nonumber \\
 & +\frac{3\sqrt{2}}{4}(\mu_{1}+\mu_{3})v_{s}+\frac{\sqrt{2}}{2}\frac{\mu^{3}}{v_{s}}\,.
\end{align}
Substituting them back in the potential allows us to compute the mass
term of the fields. When all couplings and VEVs are real, $\theta$
does not mix with the other fields, and its mass is given by
\begin{equation}
\mdmsq=-2\mu_{S}^{2}-\frac{\sqrt{2}}{2}\frac{\mu^{3}}{v_{s}}-(\lambda_{2}+4\lambda_{4})v_{s}^{2}-\lambda_{H2}v^{2}-\frac{v_{s}}{2\sqrt{2}}(\mu_{1}+9\mu_{3})-\mu_{H1}\frac{v}{2\sqrt{2}}\left(\frac{v}{v_{s}}\right)\,,\label{eq:mtheta_general}
\end{equation}
which displays the PNGB nature of the DM candidate
$\theta$, namely, its mass is zero if all the 8 symmetry breaking
couplings vanish.

On the other hand, the real parts of the fields, $(h^{\prime},\rho^{\prime})$,
do mix with a mass matrix given by
\begin{equation}
M_{S}^{2}=\begin{pmatrix}\left(M_{S}^{2}\right)_{11} & \left(M_{S}^{2}\right)_{12}\\
\left(M_{S}^{2}\right)_{21} & \left(M_{S}^{2}\right)_{22}
\end{pmatrix}\,,
\end{equation}
with the following matrix elements
\begin{align}
\left(M_{S}^{2}\right)_{11} & =2\lambda_{H}v^{2}\,,\quad\left(M_{S}^{2}\right)_{12}=\left(M_{S}^{2}\right)_{21}=\left(\lambda_{H2}+\lambda_{HS}\right)\vs v+\frac{\mu_{H1}v}{\sqrt{2}}\,,\\
\left(M_{S}^{2}\right)_{22} & =2\left(\lambda_{2}+\lambda_{S}+\lambda_{4}\right)\vs^{2}+\frac{3\vs}{2\sqrt{2}}\left(\mu_{1}+\mu_{3}\right)-\frac{\mu_{H1}v}{2\sqrt{2}}\left(\frac{v}{\vs}\right)-\frac{\sqrt{2}\mu^{3}}{2\vs}\,.
\end{align}
This mass matrix can be diagonalized by an orthogonal rotation of
angle $\alpha$
\begin{equation}
\begin{pmatrix}h\\
\rho
\end{pmatrix}=R\begin{pmatrix}h^{\prime}\\
\rho^{\prime}
\end{pmatrix}\:,\quad R\equiv\begin{pmatrix}c_{\alpha} & -s_{\alpha}\\
s_{\alpha} & c_{\alpha}
\end{pmatrix}\;,\quad RM_{S}^{2}R^{T}=\begin{pmatrix}m_{h}^{2} & 0\\
0 & m_{\text{\ensuremath{\rho}}}^{2}
\end{pmatrix}\,,
\end{equation}
where we have defined $\sa\equiv\sin\alpha$ and $\ca\equiv\cos\alpha$
in order to simplify the notation. The eigenstate $h$ is chosen to
be the $125\,\mathrm{GeV}$ boson observed at the LHC. In the numerical
analysis, we take the mixing $s_{\alpha}<0.1$ in order to satisfy
experimental measurements of the Higgs signal strengths.

One can trade-off some of the couplings in terms of the physical masses
$m_{h},m_{\rho}$ and the mixing angle $\alpha$
\begin{align}
\lambda_{H} & =\frac{c_{\alpha}^{2}m_{h}^{2}+s_{\alpha}^{2}m_{\rho}^{2}}{2v^{2}}\,,\label{eq:lambdaH}\\
\lambda_{S} & =\frac{s_{\alpha}^{2}m_{h}^{2}+c_{\alpha}^{2}m_{\rho}^{2}}{2v_{s}^{2}}-\frac{3}{4\sqrt{2}}\frac{(\mu_{1}+\mu_{3})}{v_{s}}-(\lambda_{2}+\lambda_{4})+\frac{\mu_{H1}}{4\sqrt{2}v_{s}}\left(\frac{v}{v_{s}}\right)^{2}+\frac{\sqrt{2}}{4}\frac{\mu^{3}}{v_{s}^{3}}\,,\label{eq:lambdaS}\\
\lambda_{HS} & =\frac{s_{\alpha}c_{\alpha}(m_{\rho}^{2}-m_{h}^{2})}{vv_{s}}-\frac{\mu_{H1}}{\sqrt{2}v_{s}}-\lambda_{H2}\,.\label{eq:lambdaHS}
\end{align}
Therefore, we can replace the 5 parameters of the symmetry preserving
$U(1)$ potential
\begin{equation}
m_{H},\,m_{S},\,\lambda_{H},\,\lambda_{S},\,\lambda_{HS}\,,
\end{equation}
by the 5 physical parameters
\begin{equation}
v,\,m_{h},\,v_{s},\,m_{\rho},\,s_{\alpha}\,.\label{eq:parphys}
\end{equation}
Since the mass (squared) of the DM depends linearly on all symmetry
breaking couplings one can replace one of them by the DM mass. In
the simplified cases we will consider in the next section there is
only one symmetry breaking term, therefore they will depend on 6 parameters,
those in Eq.~\eqref{eq:parphys} plus the DM mass, $\mdm$, from
which four are unknown.

\subsection{\textit{Minimal models}\label{subsec:minimal-models}}

We introduce now \textit{minimal models}, that is, with just one symmetry
breaking term. In this case the coupling can be taken real, and therefore
the DCP that stabilizes the pseudo-scalar is preserved. The selection
is motivated by either being the softest symmetry breaking term and/or
by preserving a discrete symmetry. In $V_{1,\mathrm{full}}$ and $V_{Z_{2},\mathrm{full}}$
we just choose the softest term, that is,
\begin{align}
V_{1} & =\frac{1}{2}\mu^{3}S+\mathrm{H.c.}\,,\label{eq:complex-scalar-potential-V1-1}\\
V_{Z_{2}} & =\frac{1}{2}\mu_{S}^{2}S^{2}+\mathrm{H.c.}\,.\label{eq:complex-scalar-potential-V2-1}
\end{align}
The potentials $V_{Z_{3}}$ and $V_{Z_{4}}$ in Eqs.~\eqref{eq:complex-scalar-potential-V3}
and \eqref{eq:complex-scalar-potential-V4} already contain only one
term. At some point in the text we will refer to these minimal scenarios
involving $V_{1},\,V_{Z_{2}},\,V_{Z_{3}}$ and $V_{Z_{4}}$ as linear,
quadratic, cubic and quartic models respectively.

In the numerical studies we impose the relevant theoretical constraints
on the models: perturbativity, stability of the potential, and we
have checked numerically that the minimum $(v\neq0,v_{s}\neq0)$ is
the global minimum of the potential \citep{Alanne:2020jwx}.

For the \textit{minimal models}, we particularize the expressions
for $\lambda_{H},\,\lambda_{S}$ and $\lambda_{HS}$ using the general
Eqs.~(\ref{eq:lambdaH}--\ref{eq:lambdaHS}),
\begin{equation}
\lambda_{H}=\frac{c_{\alpha}^{2}m_{h}^{2}+s_{\alpha}^{2}m_{\rho}^{2}}{2v^{2}}\,,\quad\lambda_{HS}=\frac{s_{\alpha}c_{\alpha}(m_{\rho}^{2}-m_{h}^{2})}{vv_{s}}\,,\quad\lambda_{S}=\frac{1}{2v_{s}^{2}}\left(s_{\alpha}^{2}m_{h}^{2}+c_{\alpha}^{2}m_{\rho}^{2}+A\,\mdmsq\right)\,,\label{eq:lambdas-interms-ofmasses}
\end{equation}
with $A=-1,\,0,\,1/3,\,1/2$ for the linear, quadratic, cubic and
quartic models respectively.

As can be seen in Eq.~\eqref{eq:lambdas-interms-ofmasses}, the condition
$\lambda_{S}>0$ is satisfied automatically in the \textit{minimal
models} except in the linear one, for which for the mixing angles
we will consider, $\sa\,\epsilon\,[10^{-5},\,10^{-1}]$, it just reads
$\mdmsq\lesssim m_{\rho}^{2}$.

\subsubsection{Radiative corrections}

In principle the presence of a single explicit symmetry breaking term
in the \textit{minimal models} may induce radiatively some other breaking
terms. For instance, let us consider the case of the quadratic model,
with the potential in Eq.~\eqref{eq:complex-scalar-potential-V2-1}.
At one loop it generates finite contributions to all couplings in
$V_{Z_{2},\mathrm{full}}$ (Eq.~\eqref{eq:complex-scalar-potential-V2})
and in $V_{Z_{4}}$ (Eq.~\eqref{eq:complex-scalar-potential-V4}).
For $m_{\rho}\gg m_{h}$, these corrections are of the order \footnote{DCP is still preserved in the parameter space considered if we include
these radiatively-generated terms. See Sec.~\ref{sec:beyond-minimal}.}
\begin{equation}
\lambda_{2}^{(1)}\simeq\frac{\lambda_{S}^{2}}{(4\pi)^{2}}\frac{\mu_{S}^{2}}{m_{\rho}^{2}}\,,\quad\lambda_{H2}^{(1)}\simeq\frac{\lambda_{HS}\lambda_{S}}{(4\pi)^{2}}\frac{\mu_{S}^{2}}{m_{\rho}^{2}}\,,\quad\lambda_{4}^{(1)}\simeq\frac{\lambda_{S}^{2}}{(4\pi)^{2}}\frac{\mu_{S}^{4}}{m_{\rho}^{4}}\,.
\end{equation}

Notice that $\lambda_{4}^{(1)}$ is parametrically suppressed by $\mu_{S}^{2}/m_{\rho}^{2}$
compared to $\lambda_{2}^{(1)},\lambda_{H2}^{(1)}$. This is also
the case if the couplings come from higher-dimensional operators with
spurions~\citep{Lebedev:2021xey}. We have checked that the inclusion
of the generated $\lambda_{4}^{(1)},\lambda_{H2}^{(1)}$ and $\lambda_{2}^{(1)}$
do not modify the results significantly, since their contributions
to the relevant observables are too small. 

In the other \textit{minimal models} (linear, cubic and quartic),
the symmetry breaking terms do not generate any further ones. For
the linear case this can be seen in the effective potential (i.e.,
it is not an interaction), while the cubic and quartic models in Eqs.~\eqref{eq:complex-scalar-potential-V3}
and \eqref{eq:complex-scalar-potential-V4} involve the only interactions
allowed by a $Z_{3}$ and $Z_{4}$ symmetry, respectively, and therefore
do not generate any other terms.

\section{Phenomenology\label{sec:comparison-of-models}}

\subsection{Dark matter relic abundance\label{subsec:Dark-matter-relic}}

\begin{figure}[H]
\begin{centering}
\subfloat[]{\begin{centering}
\includegraphics[scale=0.4]{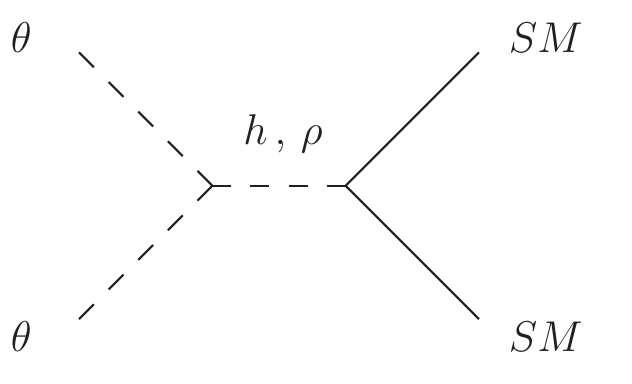}
\par\end{centering}
}
\par\end{centering}
\begin{centering}
\subfloat[]{\begin{centering}
\includegraphics[scale=0.4]{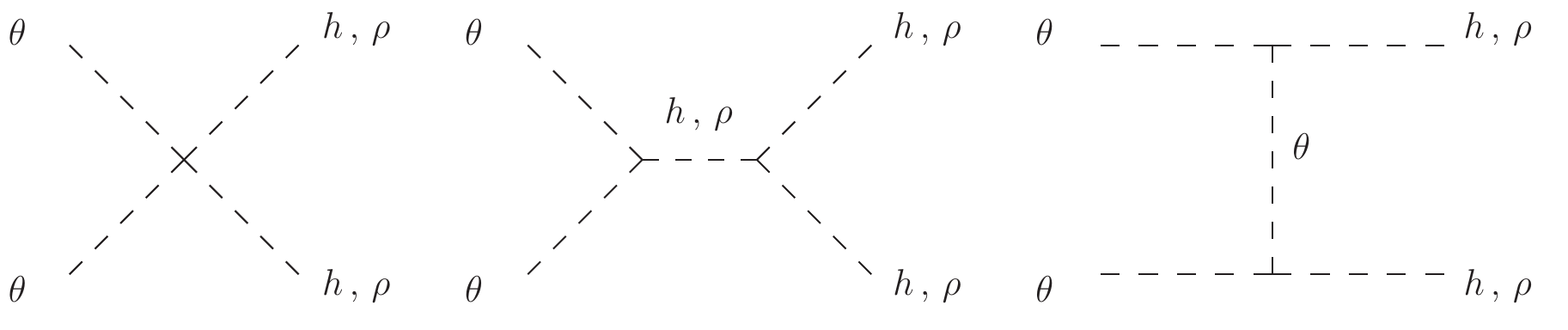}
\par\end{centering}
}
\par\end{centering}
\caption{\label{fig:Feyndiagrams}Feynman diagrams of the possible DM annihilation
channels. (a) DM annihilations into SM states mediated by the Higgs
boson $h$ and the scalar $\rho$. (b) DM annihilations into $h,\rho$
for the case of forbidden and secluded DM.}
\end{figure}

In the Early Universe, the complex scalar singlet is in thermal equilibrium
with the SM thanks to the interactions mediated by the Higgs portal
coupling, $\lambda_{HS}$. Then, at some point in the evolution of
the Universe, the scalar $S$ acquires a real VEV, breaking all possible
symmetries except DCP, which, after SSB, is manifested by the pseudo-scalar
$\theta$ as $\theta\rightarrow-\theta$, which could therefore be
a potential DM candidate. Once its interactions are slow enough compared
to the expansion of the Universe, it freezes-out and thereafter its
number density over entropy density, $n/s$, remains constant. Considering
non-relativistic freeze-out, in large regions of the parameter space
the interactions are too weak, the pseudo-scalar freezes-out too early,
and the relic abundance is too large. However, there are parts of
the parameter space that may produce large enough annihilations, so
that the correct relic abundance is reproduced:
\begin{enumerate}
\item Resonances with the Higgs boson $h$ or with the scalar $\rho$, which
occur for $2m_{\theta}\text{\ensuremath{\simeq}}m_{h}$ or $2m_{\theta}\text{\ensuremath{\simeq}}m_{\rho}$
(See Fig.~\ref{fig:Feyndiagrams} (a)). In this region, kinetic equilibrium
of the final states at freeze-out may not be a good assumption, and
therefore there is some uncertainty in the parameter values for which
the abundance is reproduced \citep{Binder:2017rgn,Binder:2021bmg,Abe:2021jcz}.
\item Direct annihilations into (lighter) pairs of scalars $h$ and/or $\rho$,
for $m_{\theta}\text{\ensuremath{\gtrsim}}m_{h}$ and/or $m_{\theta}\gtrsim m_{\rho}$
(See Fig.~\ref{fig:Feyndiagrams} (b)). The latter case is known
as secluded dark matter (SDM). If $\lambda_{HS}\neq0$, for $s_{\alpha}\gtrsim10^{-16}$,
$\rho$ decays via mixing into SM states, as long as $m_{\rho}>2m_{e}$.

\item Direct annihilations into somewhat slightly heavier pairs of $hh,h\rho,\rho\rho$
(See Fig.~\ref{fig:Feyndiagrams} (b)). This is known as forbidden
dark matter (FDM). For $m_{h}\gtrsim m_{\rho}\gtrsim m_{\theta}$,
annihilations into $\rho\rho$ are normally larger and set the abundance.
For $m_{\theta}\lesssim m_{h}\lesssim m_{\rho}$, which channel dominates
depends on the mixing angle $\alpha$. In Ref.~\citep{Binder:2021bmg} the case in which kinetic equilibrium of the final states at freeze-out may not be a good assumption has also been studied.
\item Non-resonant Higgs-mediated annihilations into SM states happening
for DM masses above 100 GeV and at mixings $s_{\alpha}$ larger than
the previous cases.~\citep{Arina:2019tib,Azevedo:2018oxv} (See Fig.~\ref{fig:Feyndiagrams}
(a)).
\end{enumerate}
Along this paper we use the DM relic abundance value from the Planck
Collaboration \citep{Aghanim:2018eyx}
\begin{equation}
\Omega_{\text{\text{obs}}}h^{2}=0.120\pm0.001\,,\label{eq:relic_abundance}
\end{equation}
where $h$ is the dimensionless Hubble parameter, $H=100\,h\,$ km/s/Mpc.

\subsection{Direct detection}

When there is non-zero mixing between the CP-even scalars, the pseudo-scalar
may give rise to nuclear scatterings in underground detectors. In
the limit of an exact Goldstone boson, the scatterings of $\theta$
are suppressed by the small momentum transfer. However, for the \textit{minimal
models} with different symmetry breaking terms, in principle significant
interaction rates may be generated, which may allow one to distinguish
them. We follow the analysis of Refs. \citep{Alanne:2020jwx,Gross:2017dan}.
The spin-independent DD cross section at tree level is given by
\begin{equation}
\frac{d\sigma_{SI}}{d\Omega}=\frac{\lambda_{SI}^{2}\,f_{N}^{2}\,m_{N}^{2}}{16\pi^{2}\,m_{\theta}^{2}}\left(\frac{m_{\theta}m_{N}}{m_{\theta}+m_{N}}\right)^{2}\,,\label{eq:dsigmaSIDD}
\end{equation}
where $m_{N}=0.939\,\gev$ is the nucleon mass, and $f_{N}=0.28(7)$ is
the effective Higgs-nucleon coupling \citep{Alarcon:2011zs,Alarcon:2012nr}. The effective
DM-nucleon coupling $\lambda_{SI}$ reads as
\begin{equation}
\lambda_{SI}^{2}\equiv\frac{1}{4f_{N}^{2}m_{N}^{4}}|\mathcal{M}|^{2}=\frac{1}{4m_{N}^{2}v^{2}}\left(\frac{\beta_{h\theta\theta}\,c_{\alpha}}{t-m_{h}^{2}}+\frac{\beta_{\rho\theta\theta}\,s_{\alpha}}{t-m_{\rho}^{2}}\right)^{2}(4m_{N}^{2}-t)\,,\label{eq:lambSIdefinition}
\end{equation}
and the $\beta_{i}$ coefficients are defined by the interactions
described in the Appendix~\ref{sec:beta-coeff}. One can see that
in the zero momentum limit $(t\rightarrow0)$, the effective DM-nucleon
coupling goes as
\begin{equation}
\lambda_{SI}\propto-\left(\frac{\beta_{h\theta\theta}\,c_{\alpha}}{m_{h}^{2}}+\frac{\beta_{\rho\theta\theta}\,s_{\alpha}}{m_{\rho}^{2}}\right)\,.\label{eq:lambSI}
\end{equation}
The expressions for the effective coupling in the small momentum limit
are particularized for the \textit{minimal models} in Table \ref{tab:effDMnucleoncoupling}.
As the coupling $\lambda_{SI}$ enters squared in the cross section,
there will be no difference for the linear and cubic models, whereas
in the case of the quartic model the effective coupling squared is a factor
of 4 larger with respect to the former models. For the quadratic model
the effective DM-nucleon coupling vanishes in the zero-momentum limit
\citep{Gross:2017dan,Alanne:2020jwx}. We will show this using an
EFT approach in Sec.~\ref{sec:pGB-limit}.
\begin{table}[H]
\begin{centering}
\begin{tabular}{|c|c|}
\hline 
\textit{Minimal model} & $\lambda_{SI}\propto-\left(\frac{\beta_{h\theta\theta}\,c_{\alpha}}{m_{h}^{2}}+\frac{\beta_{\rho\theta\theta}\,s_{\alpha}}{m_{\rho}^{2}}\right)$\tabularnewline
\hline 
\hline 
Linear & $\frac{s_{\alpha}c_{\alpha}}{\vs m_{h}^{2}m_{\rho}^{2}}m_{\theta}^{2}(m_{h}^{2}-m_{\rho}^{2})$\tabularnewline
\hline 
Quadratic & $0$\tabularnewline
\hline 
Cubic & $-\frac{s_{\alpha}c_{\alpha}}{\vs m_{h}^{2}m_{\rho}^{2}}m_{\theta}^{2}(m_{h}^{2}-m_{\rho}^{2})$\tabularnewline
\hline 
Quartic & $-2\frac{s_{\alpha}c_{\alpha}}{\vs m_{h}^{2}m_{\rho}^{2}}m_{\theta}^{2}(m_{h}^{2}-m_{\rho}^{2})$\tabularnewline
\hline 
\end{tabular}
\par\end{centering}
\caption{\label{tab:effDMnucleoncoupling}Effective DM-nucleon coupling that
enters in the DD cross section in terms of the physical parameters
$v,\,\protect\vs,\,m_{h},\,\protect\mdm,\,m_{\rho}$ and $s_{\alpha}$.}
\end{table}

We assume that the one-loop contributions are small compared to the
tree level ones, as we are not in any particular point in parameter
space where cancellations at tree level occur \citep{Alanne:2020jwx}
(see also Refs.~\citep{Abe:2021nih,Glaus:2020ihj,Ishiwata:2018sdi,Azevedo:2018exj}
for one-loop contributions in the case of the quadratic model, where
indeed they are the dominant contribution).

In the numerical analysis, DD constraint from XENON1T \citep{Aprile:2018dbl}
has been taken into account using the relative DM relic abundance
of each point in the scan, and rescaling the experimental bound. We
are therefore assuming that the local relic abundance scales like
the global one, e.g. $\rho\propto\Omega$. That is, every allowed
point in the parameter space satisfies
\begin{equation}
\frac{\Omega}{\Omega_{\text{obs}}}\sigma_{SI}\leqslant\sigma^{\text{XENON1T}}\,,\label{eq:rescaledbound}
\end{equation}
where $\sigma^{\text{XENON1T}}$ is the $90\%$ confidence level upper
limit on the DM-nucleon spin-independent cross section from XENON1T.

\subsection{Higgs signal strength and invisible decays}

For $m_{\theta}<m_{h}/2$ and/or $m_{\rho}<m_{h}/2$ new decay channels
of the Higgs boson into $\theta$ and $\rho$ open up. The decay widths
into these final states read
\begin{equation}
\Gamma(h\rightarrow\theta\theta)=\frac{\beta_{h\theta\theta}^{2}}{32\pi\,m_{h}}\sqrt{1-\frac{4m_{\theta}^{2}}{m_{h}^{2}}}\,,\qquad\Gamma(h\rightarrow\rho\rho)=\frac{\beta_{h\rho\rho}^{2}}{32\pi\,m_{h}}\sqrt{1-\frac{4m_{\rho}^{2}}{m_{h}^{2}}}\,,
\end{equation}
with the $\beta_{i}$ coefficients defined in Appendix~\ref{sec:beta-coeff}.
The Higgs invisible branching ratio is therefore given by
\begin{equation}
{\rm BR}(h\rightarrow\text{inv})=\frac{\Gamma(h\rightarrow\theta\theta)+\Gamma(h\rightarrow\rho\rho)}{c_{\alpha}^{2}\Gamma_{h}^{{\rm SM}}+\Gamma(h\rightarrow\theta\theta)+\Gamma(h\rightarrow\rho\rho)}\,,
\end{equation}
where $\Gamma_{h}^{{\rm SM}}=4.1\,\mev$ is the SM Higgs decay width.
The observed Higgs signal strength in the measured channels imposes
a constraint on $\mu=c_{\alpha}^{2}(1-{\rm BR}(h\rightarrow\text{inv}))$.
A detailed analysis taking into account the contributions of both
CP-even scalars to the different channels is beyond the scope of this
work. However, in the limit of small mixing $s_{\alpha}$ and very
heavy scalar $\rho$, we can neglect the contribution of the latter,
take $c_{\alpha}^{2}=1$ and directly impose the experimental constraints
on the invisible Higgs width. If the mass of $\rho$ is of the order
of the Higgs boson mass or below (SDM), and if the mixing is very
small, this is also a reasonable assumption. We therefore impose the
90\% confidence level upper limit of ${\rm BR}(h\rightarrow\text{inv})<0.16$
obtained by the CMS Collaboration \citep{Sirunyan:2018owy} (see also
Ref.~\citep{Aaboud:2019rtt} for the ATLAS Collaboration results). 

\subsection{Numerical results}

In the following we present the results of a numerical analysis of
the scenarios discussed so far. First, we focus on the quadratic model
as prototypical example, which also features suppressed DD rates,
as we have discussed. The goal is to obtain all the regions where
the correct relic abundance can be reproduced. Then we investigate
the parameter space where it is possible to have a good DM candidate
in each of the \textit{minimal models}. Notice that if the symmetry
breaking terms $\lambda_{H2}$ and $\mu_{H1}$ in Eqs.~\eqref{eq:complex-scalar-potential-V1}
and \eqref{eq:complex-scalar-potential-V2} vanish, all the observables
we consider depend on $s_{\alpha}^{2}$. Therefore, we will only choose
positive values for the mixing. Finally we analyze the possibility
to distinguish the models comparing their predictions to different
observables. The relic abundance has been computed using the code
\texttt{micrOMEGAs} \citep{Belanger:2018ccd}, see also Ref.~\citep{Gondolo:1990dk}.

\subsubsection{The quadratic model}

\begin{figure}[H]
\begin{centering}
\includegraphics[scale=0.3]{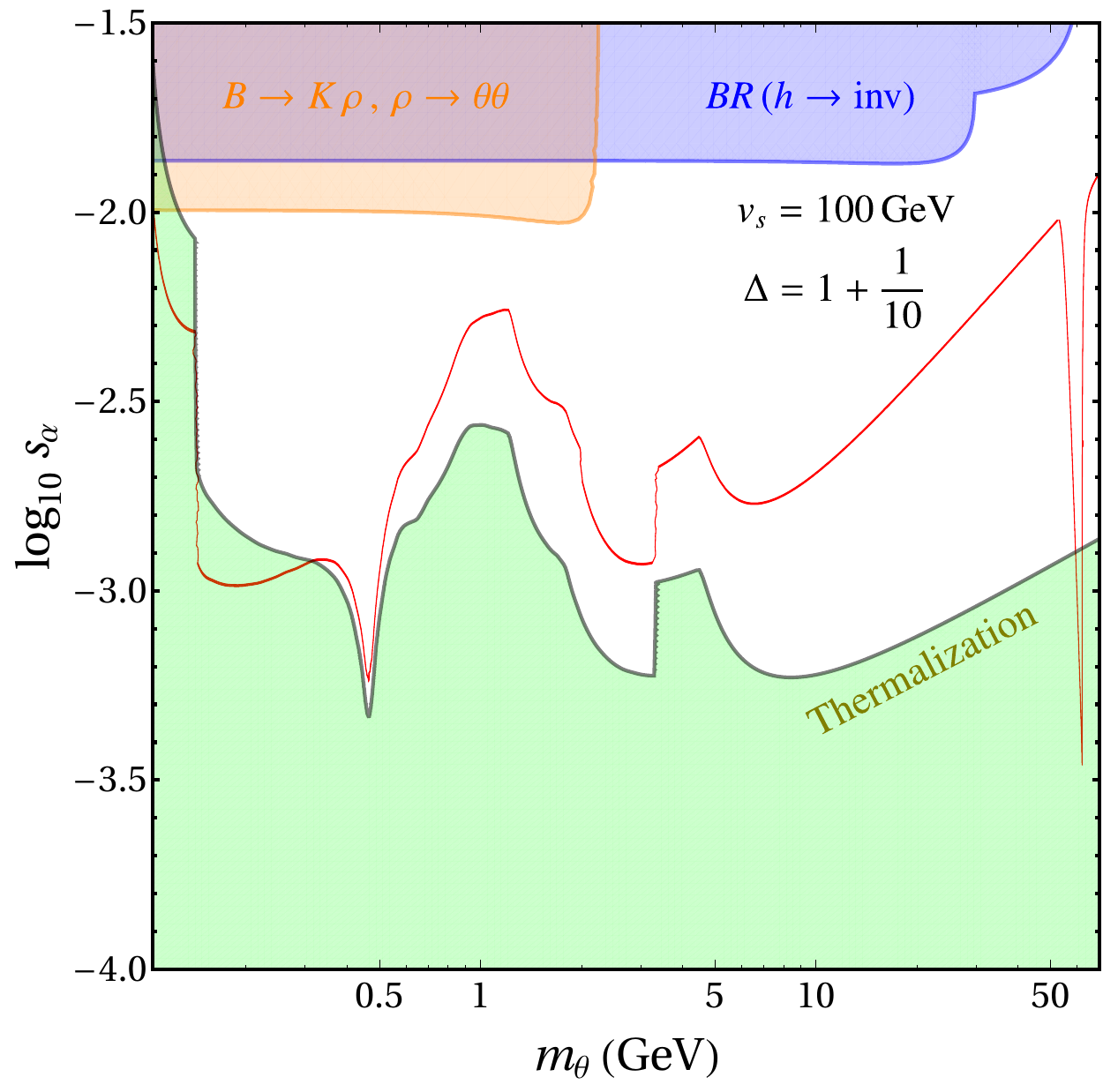}~\includegraphics[scale=0.3]{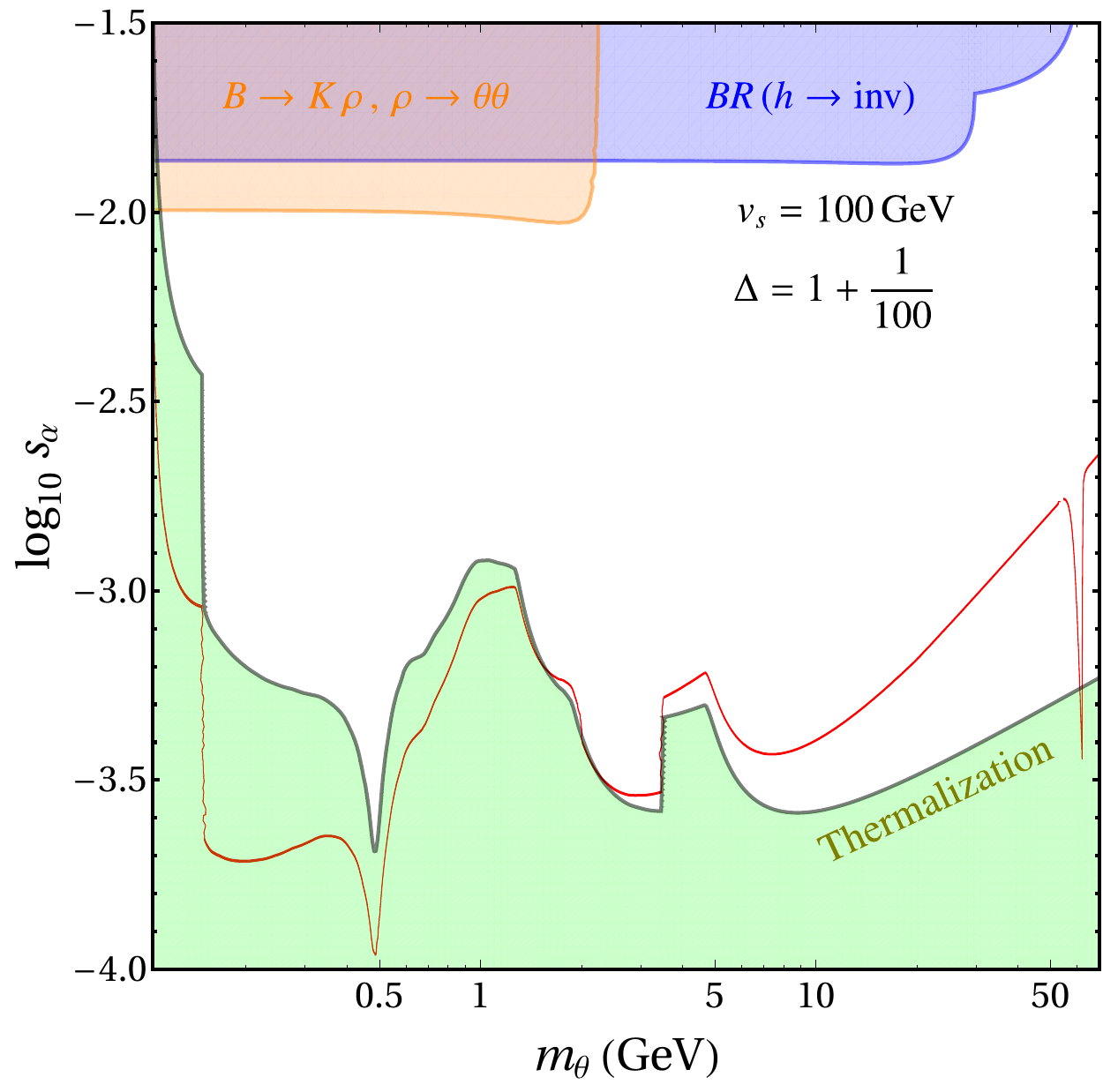}
\par\end{centering}
\caption{\label{fig:combinedplot} The red line shows the relic
abundance in Eq.~\eqref{eq:relic_abundance} for the resonance condition
with $\rho$ in the quadratic model. We plot different values of $\Delta=1+1/10,\,1+1/100\,$
 on the left and right panels respectively, see Eq.~\eqref{eq:res_condition}, with $\vs=100~\gev$. Experimental constraints from invisible Higgs
decays (blue), rare $B$ meson decays into light scalars (orange)
and the thermalization condition (green) are shown.}
\end{figure}

In this section we focus on one of the possibilities to obtain the
correct relic abundance: setting the DM candidate $\theta$ in resonance
with the scalar $\rho$, which corresponds to the option 1 discussed
in Sec.~\ref{subsec:Dark-matter-relic}. The relevant annihilation
process is $\theta\theta\rightarrow ff$, where $f$ refers to some
SM particle. This setting allows one to expand the parameter space of
DM masses, which are not necessarily needed to be close to the Higgs
boson resonance \citep{Arina:2019tib,Azevedo:2018oxv}. We parametrize
the deviation from the resonance condition with the dimensionless
mass splitting parameter

\begin{equation}
\text{\ensuremath{\Delta=\frac{(m_{\rho}-\mdm)}{\mdm}}}\,,\label{eq:res_condition}
\end{equation}
so that $m_{\text{\ensuremath{\rho}}}=\left(\Delta+1\right)m_{\theta}$.
In Fig.~\ref{fig:combinedplot} we depict the (logarithm of the)
mixing that reproduces the correct relic abundance in Eq.~\eqref{eq:relic_abundance}
as a function of the DM mass for different values of the dimensionless mass splitting $\Delta$ and an illustrative value $\vs=100~\gev$. For masses below $4$-$5\,\gev$
decays into hadrons are taken into account\footnote{Special attention is needed here as \texttt{micrOMEGAs} does not consider
hadronic final states, see details in Appendix~\ref{sec:Hadronization-effects}.} and also constraints from having a light scalar mixing with the Higgs
boson \citep{Winkler:2018qyg}. In the parameter space considered
here, the most relevant constraints come from the invisible
Higgs boson decays, the limits on rare $B$-meson decays, and being in thermal equilibrium with the SM particles in the Early Universe. The orange
shaded region is excluded by the limits set on $B\rightarrow K\rho\rightarrow K\,+\text{``invisible"}$
\citep{Zyla:2020zbs}. In our model, $\rho$ decays into an invisible
final state composed of two $\theta$, and both escape the detectors
leaving no signal. The calculation of the decay rate $B\rightarrow K\rho$
was performed using the expressions in Ref.~\citep{Winkler:2018qyg}.
The other relevant constraint is set by the invisible Higgs decay,
shown as the blue shaded region.

The features of the curve in Fig.~\ref{fig:combinedplot} satisfying the relic abundance can be understood from Fig.~4 of Ref.~\citep{Winkler:2018qyg}. Below the muon mass, DM annihilations can only occur into electrons and positrons and are very suppressed, requiring large mixing angles that are constrained. Above the muon mass, the kinks correspond to the opening of annihilation channels into different hadronic and leptonic states. Also, at $m_{\theta}\simeq60$
GeV the Higgs boson resonance is clearly visible. It is important to mention how these results change for different values of $\vs$. On one side, there is a cancellation in the annihilation rate on the resonance, making it independent of $\vs$. On the other side, the invisible Higgs limits weaken for larger values of $\vs$.

Notice that a detailed analysis of the relic abundance very close
to the resonance has been performed in Ref.~\citep{Binder:2017rgn},
and also for the case of a PNGB in Ref.~\citep{Abe:2021jcz}, where
significant differences appear in the computation of the relic density
depending on whether kinetic equilibrium is maintained at freeze-out
or not. Also, there is a huge sensitivity to the mass splitting, as
can be seen by comparing the required mixing angles for the two different
values of the parameter $\Delta$. Therefore, Fig.~\ref{fig:combinedplot}
is intended to show that it is possible to have the proper relic abundance
in the considered parameter space, but should be taken with a grain
of salt regarding the precision of the exact value of the mixing angle
for a given mass splitting that reproduces the abundance.

\subsubsection{Comparison of \textit{minimal models }\label{subsec:Comparison-of-minimal}}

Now, we consider all the \textit{minimal models} and study their allowed
parameter space.

\begin{figure}[H]
\begin{centering}
\includegraphics[scale=0.3]{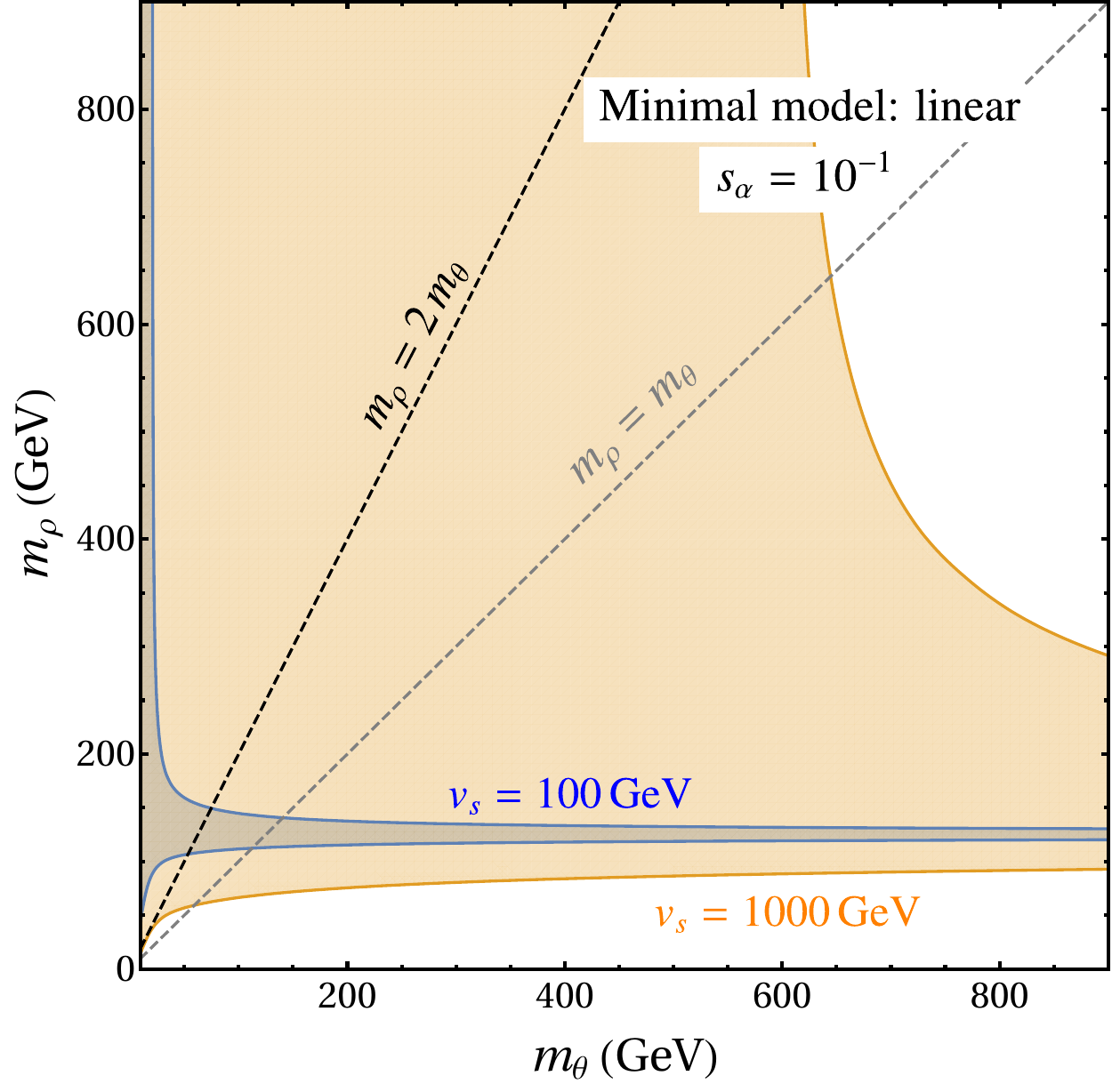}~\includegraphics[scale=0.3]{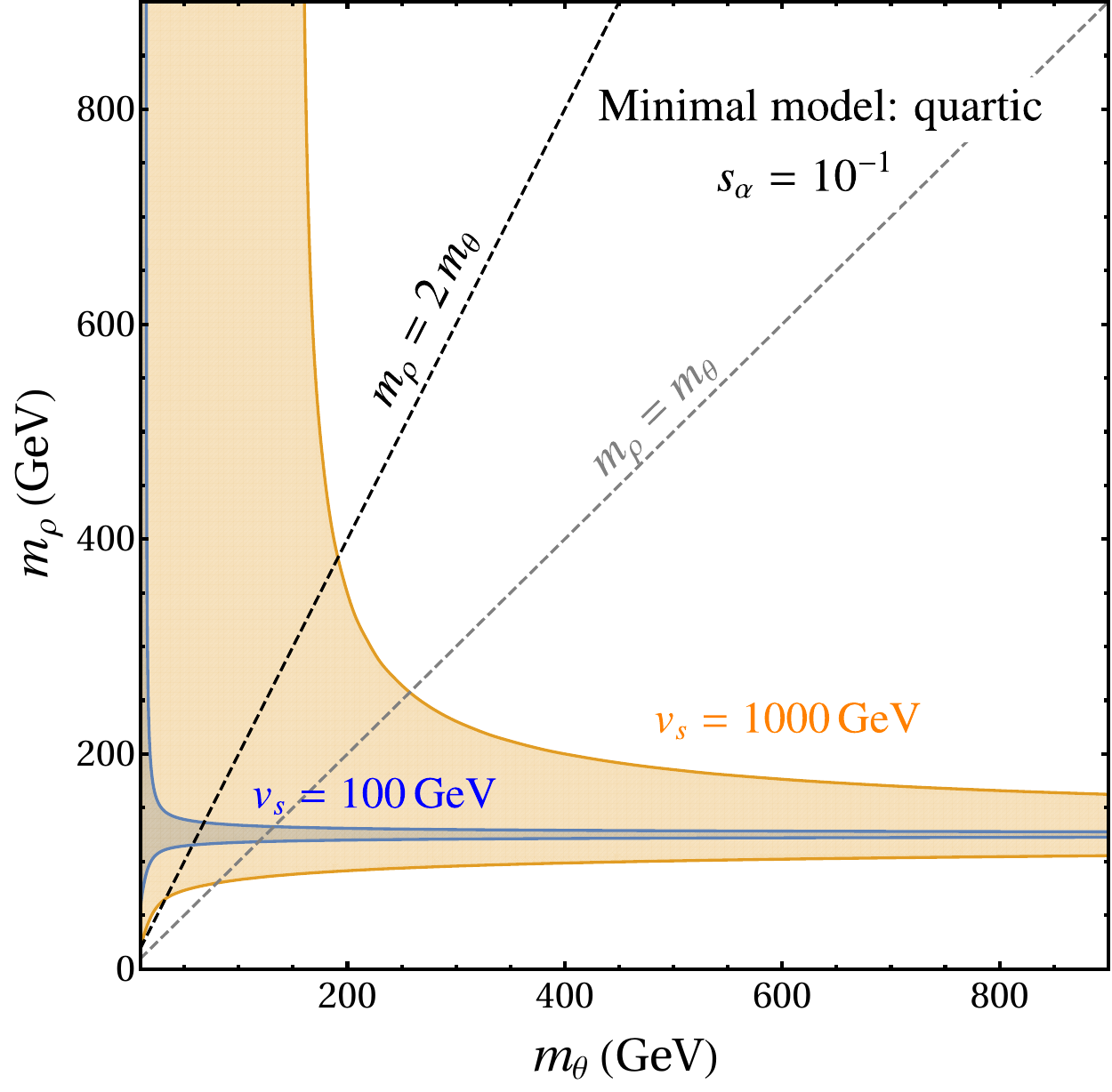}
\par\end{centering}
\caption{\label{fig:DDregplotmodels}The colored region is the parameter space
allowed by the XENON1T null-results for the linear (left) and quartic
(right) models. Blue (Orange) colored region corresponds to $\protect\vs=100\,(1000)\,\protect\gev$.
Black and gray dashed lines represent the resonance condition with
the $\rho$ $(m_{\rho}=2m_{\theta})$ and the degenerate case $(m_{\rho}=m_{\theta})$
respectively.}
\end{figure}

First, we focus on the restrictions that DD limits set on the \textit{minimal
models}, irrespectively of the relic abundance. In Fig.~\ref{fig:DDregplotmodels}
we plot the allowed parameter space for the different models in the
$\left(\mdm,m_{\rho}\right)$ plane for $\sa=10^{-1}$ for two values
of $v_{s}$ ($100$ GeV in blue, $1$ TeV in orange), after imposing
the XENON1T limits \citep{Aprile:2018dbl}. In order to do that, we
use the expression for the DD cross section in Eq.~\eqref{eq:dsigmaSIDD}
and the $\lambda_{SI}$ values shown in Table~\ref{tab:effDMnucleoncoupling}.
The quadratic model has a momentum-suppressed DD cross section at
tree level and therefore it is not shown in the plot. Moreover, the
effective coupling $\lambda_{SI}$ for the cubic model is similar
to the linear case, apart from a relative sign, so we only plot the
results for the latter one. For smaller values of the mixing, there
are no restrictions from DD null-results, except for very light $\rho$
masses. This can be understood by looking at the expressions of the
effective coupling in Table~\ref{tab:effDMnucleoncoupling} and its
dependence on the $\rho$ mass, which goes as $\lambda_{SI}\simeq1/m_{\rho}^{2}$
for small $m_{\rho}$ values (but still larger than the typical momentum
transfer at the XENON1T Experiment, $\mathcal{O}(\mev$)).

In Appendix~\ref{sec:cases-H1-H2} we also discuss DD constraints
for the symmetry breaking terms involving the Higgs field in Eqs.~\eqref{eq:complex-scalar-potential-V1}
and \eqref{eq:complex-scalar-potential-V2}, which are qualitatively
different.

Now we impose the requirement of reproducing the relic abundance. First of all, we have checked that in all cases DM is in thermal equilibrium with SM particles in the Early Universe: in the resonance case the relevant process is $\theta \theta \leftrightarrow {\rm SM}\, {\rm SM}$, and in the SDM/FDM scenarios the relevant ones are $\theta \theta \leftrightarrow \rho \rho, h \rho, hh$ followed by $\rho  \leftrightarrow {\rm SM}\, {\rm SM}$. Therefore, we can safely use \texttt{micrOMEGAs} \citep{Belanger:2018ccd}
for the numerical computation. Moreover, in the case of SDM/FDM, we have checked that for the masses and mixings considered, the $\rho$ particles
always decay into SM states. 

\begin{figure}[H]
\begin{centering}
\includegraphics[scale=0.26]{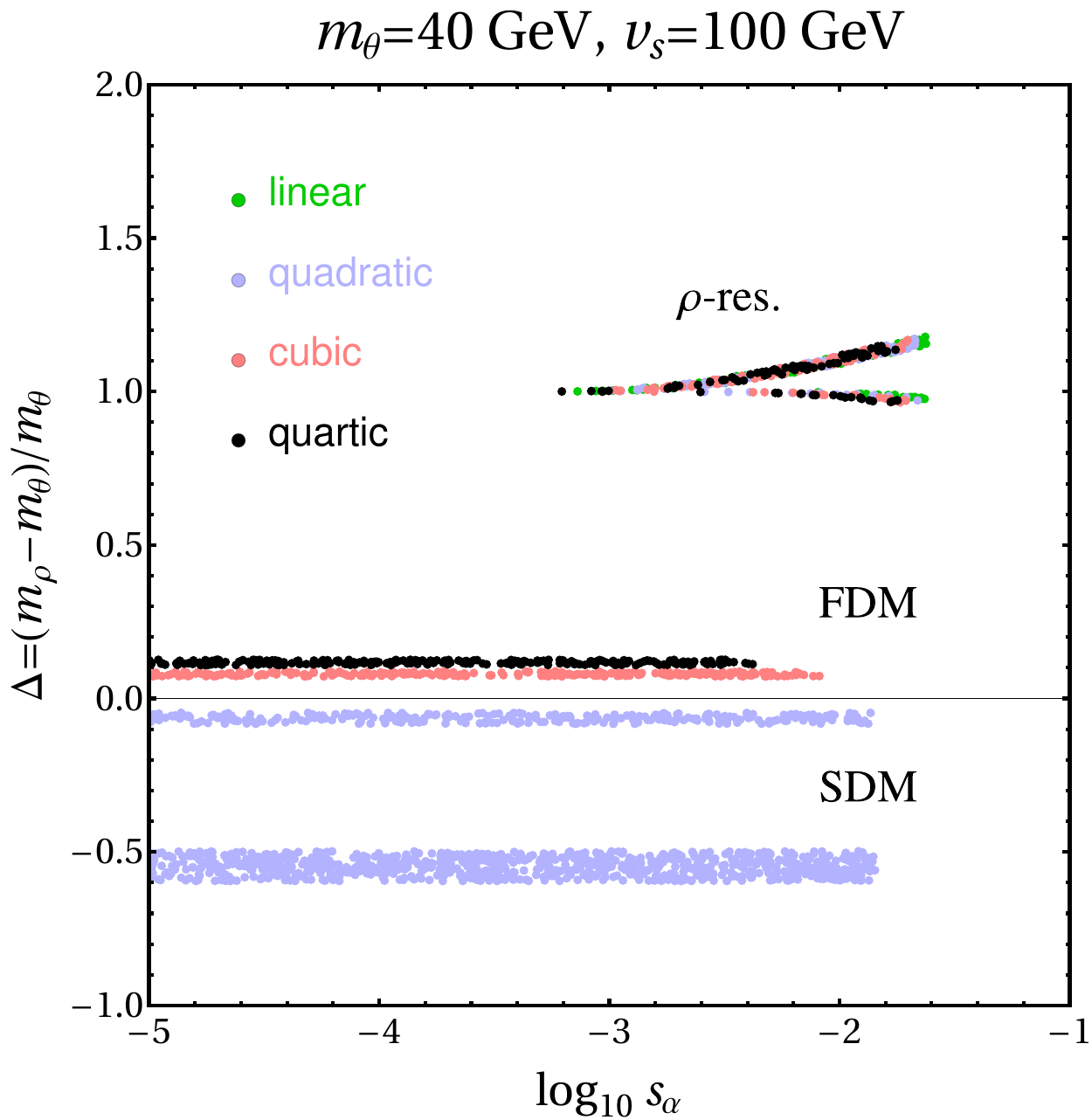}\includegraphics[scale=0.26]{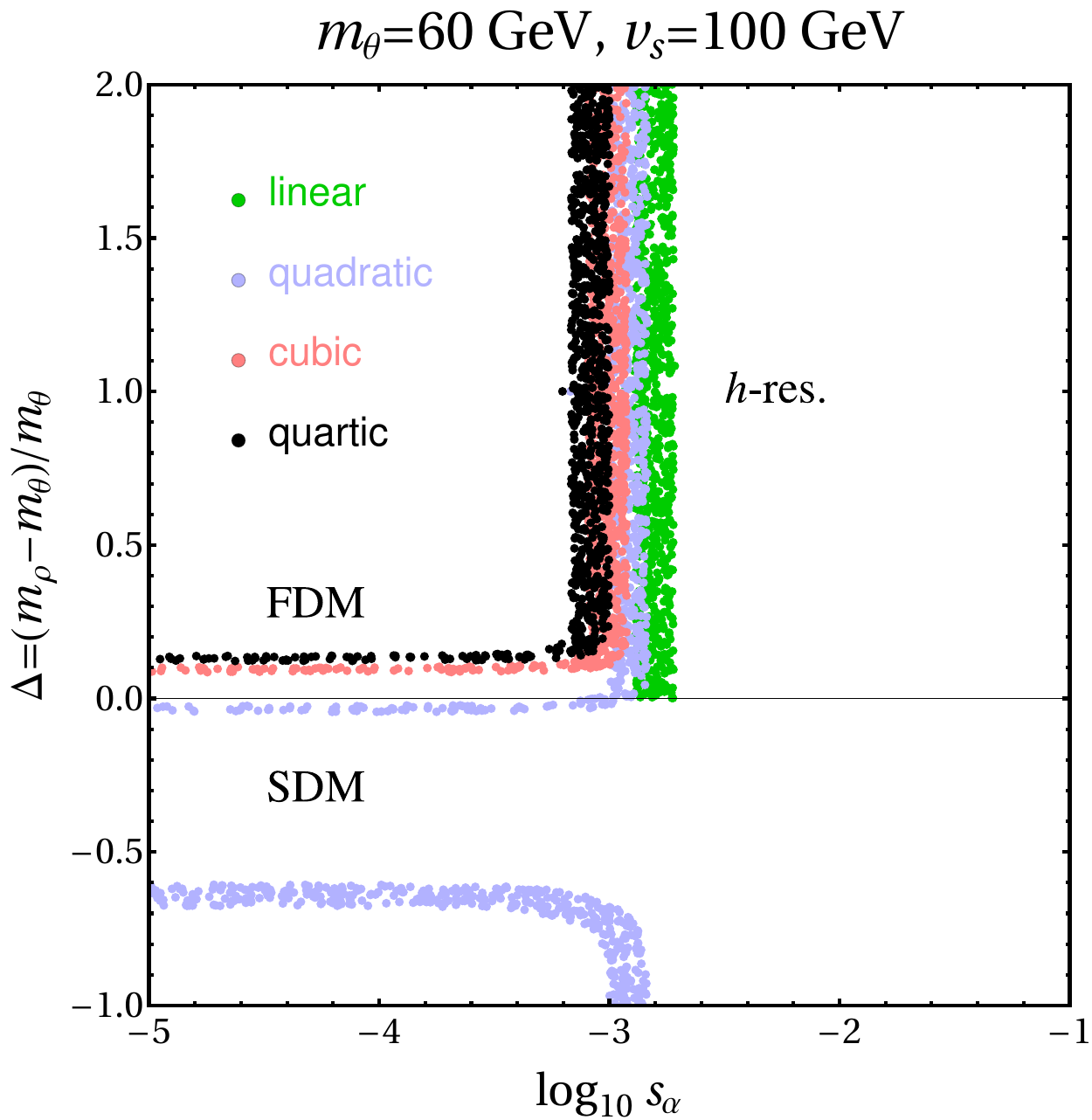}\includegraphics[scale=0.26]{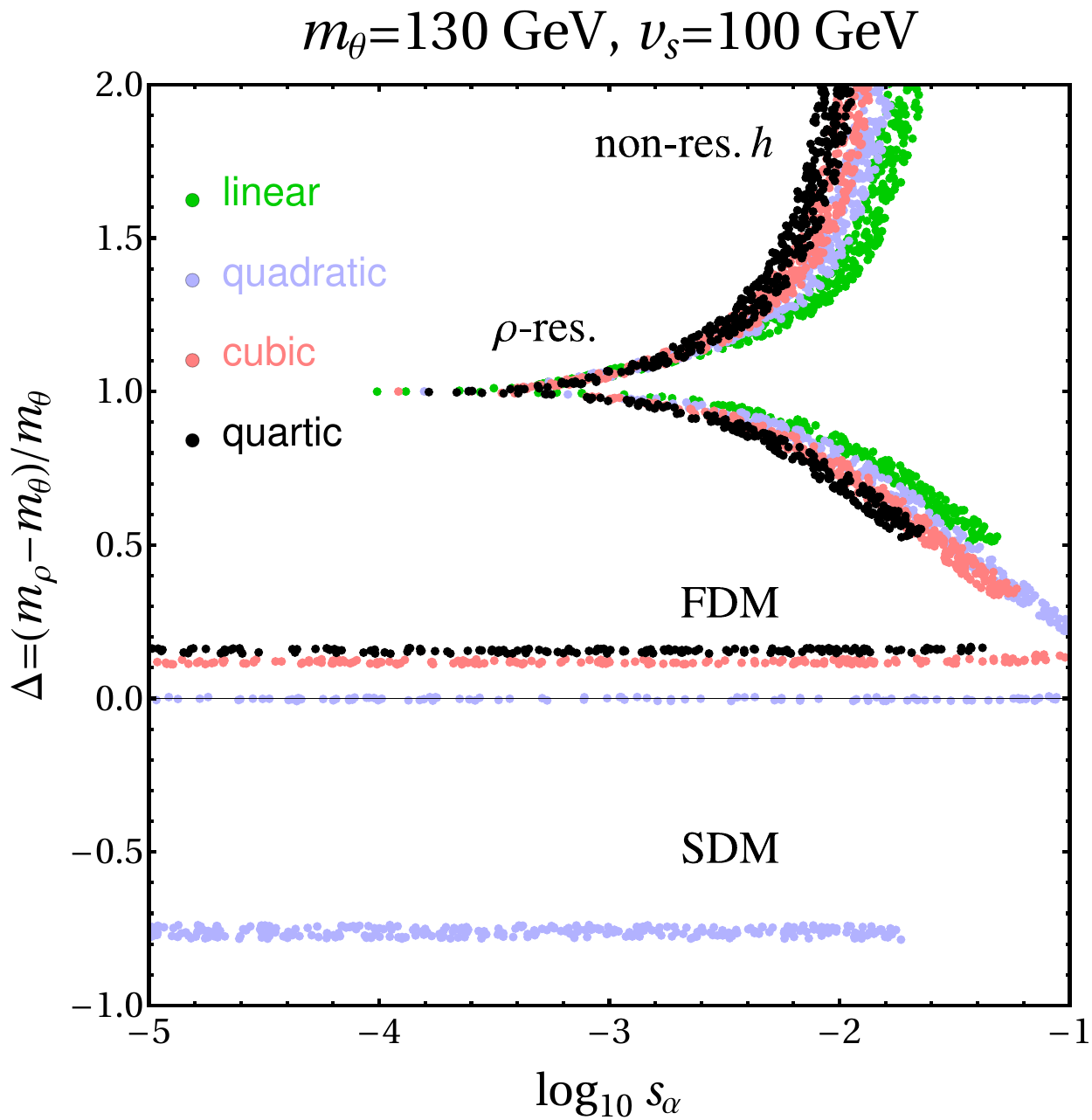}
\par\end{centering}
\caption{\label{fig:Scandeltamixing}Scan in the normalized mass splitting
$\Delta$ versus the (logarithm of the) mixing $\protect\sa$ for
$\protect\mdm=40,\,60,\,130\,\protect\gev$ from left to right, and
$\protect\vs=100\,\protect\gev$, for the \textit{minimal models}
described in the text. The green, blue, red and black colors correspond
to the linear, quadratic, cubic and quartic models respectively. All
these points fulfill the relic abundance condition $0.5\leqslant\Omega/\Omega_{\text{obs}}\leqslant1$.
We also impose the XENON1T and the invisible Higgs decay constraints.}
\end{figure}

In Fig.~\ref{fig:Scandeltamixing} we show the results of a scan
of the normalized mass splitting parameter $\Delta$ and the mixing
angle for three different values of the mass (for $\mdm=40,\,60,\,130\,\gev$
from left to right) and a fixed value of the VEV ($v_{s}=100$ GeV),
where all the points satisfy that $0.5\leqslant\Omega/\Omega_{\text{obs}}\leqslant1$.
As can be observed, the different freeze-out scenarios discussed in
Sec.~\ref{subsec:Dark-matter-relic} are realized in separated regions:
the Higgs resonance ($h$-res.) for $m_{\theta}\simeq m_{h}/2$ at
a fixed value of the mixing, the $\rho$ resonance ($\rho$-res.)
for $\Delta\simeq1$, FDM when $\Delta\gtrsim0$, SDM for $\Delta<0$
and the non-resonant Higgs-mediated annihilations (non-res. $h$)
for masses above 100 GeV and $s_{\alpha}$ larger than the previous
cases. It is important to point out
that the Higgs resonance is only observed in the middle plot and that
is in agreement with Fig.~\ref{fig:combinedplot}, where for $m_{\theta}\simeq m_{h}/2$
the main DM annihilation channel into SM particles is mediated by
the Higgs boson. As expected, the resonance regions are the ones changing
the most among the different plots. Note that for the linear model,
the proper relic abundance is reached only at the resonances. This
is due to the fact that the theoretical constraint $\lambda_{S}>0$
restricts the values of the mass splitting to the region $\Delta>0$.
Therefore, SDM can not be realized in this case. Notice also that
the perturbativity constraint ($\lambda_{S}<4\pi$) reduces the parameter
space of each model in the case of large $\rho$ masses. 

In Ref.~\citep{Azevedo:2018oxv} similar results where obtained for
the case in which the relic abundance is reproduced through processes
via resonances, and for the cases of large mixing and masses above
$100\,\gev$. The former can be seen for all $\mdm$ masses considered
in Fig.~\ref{fig:Scandeltamixing}, and the latter in the case of
$\mdm=130\:\gev$, where the resonant region starts expanding from
$\Delta\simeq1$ (non-res. $h$). However the regions with proper
relic values from FDM and SDM scenarios were not discussed in that
work.
\begin{figure}[H]
\begin{centering}
\includegraphics[scale=0.35]{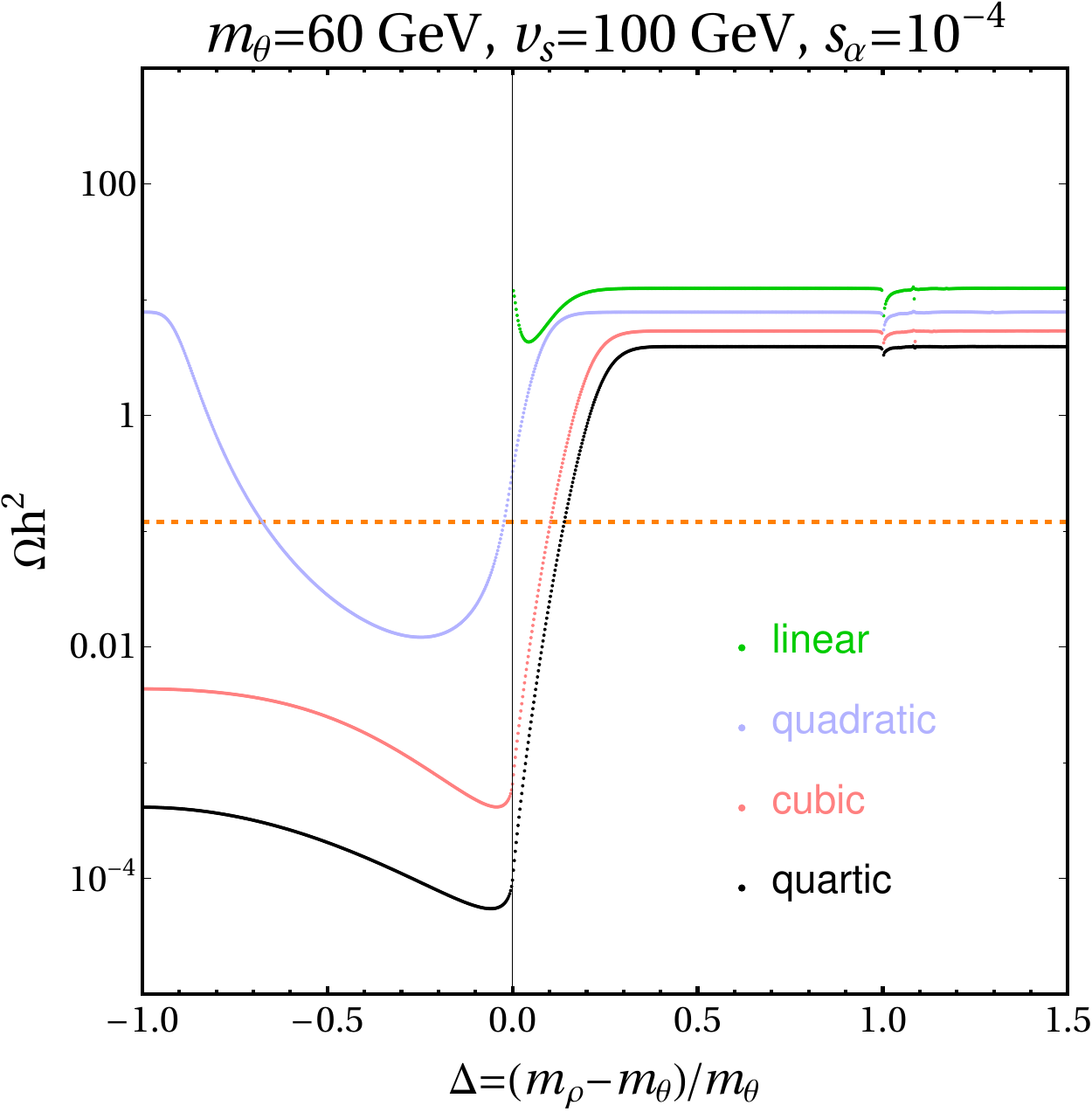}
\par\end{centering}
\caption{\label{fig:m60v100relicdelta}Relic abundance as a function of $\Delta$
for $\protect\mdm=60\,\protect\gev$, $\protect\vs=100\,\protect\gev$
and $\protect\sa=10^{-4}$ for the \textit{minimal models}. The correct
value for the relic abundance is shown as an orange dashed line.}
\end{figure}

In order to better understand the structure of strips of Fig.~\ref{fig:Scandeltamixing},
we plot in Fig.~\ref{fig:m60v100relicdelta} the relic abundance
for fixed values of $\mdm=60\,\gev$, $\vs=100\,\gev$ and $\sa=10^{-4}$
. We observe that for the case of FDM, close to $\Delta\gtrsim0$,
the linear model does not reach a cross section large enough in order
to have the correct relic abundance. This is due to a partial cancellation
happening among the diagrams (see the sign difference in Eq.~\eqref{eq:lambdaS}
in the contribution to $\lambda_{S}$ of $\mu^{3}$ with respect to
the $\mu_{3}$ or $\lambda_{4}$ one). In the quadratic model, for
$\Delta<0$ we see that SDM is allowed for two different values of
the mass splitting. The explanation comes from the fact that the amplitude
for the DM annihilations into $\rho$ has two contributions, one proportional
to $m_{\rho}$ and the other one to $m_{\theta}$, and in this model
only the former is present, so when $m_{\rho}\rightarrow0$ the amplitude
vanishes at tree level\footnote{The values of $\Omega h^{2}$ in the limit $m_{\rho}\rightarrow0$
displayed in Fig.~\ref{fig:m60v100relicdelta} should not be trusted
because the amplitude of the process goes to zero.}.

\begin{figure}[H]
\begin{centering}
\includegraphics[scale=0.3]{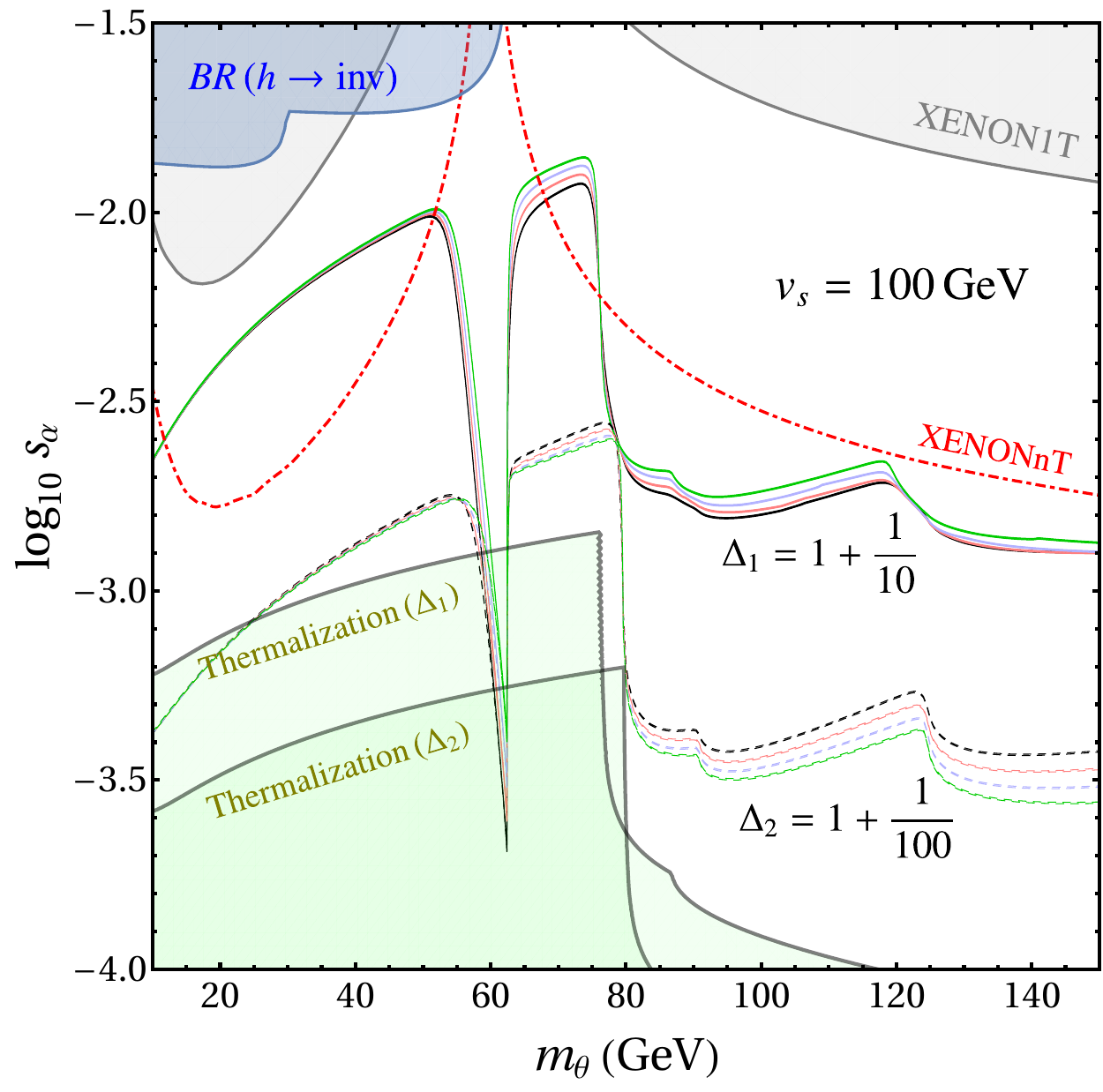}~\includegraphics[scale=0.3]{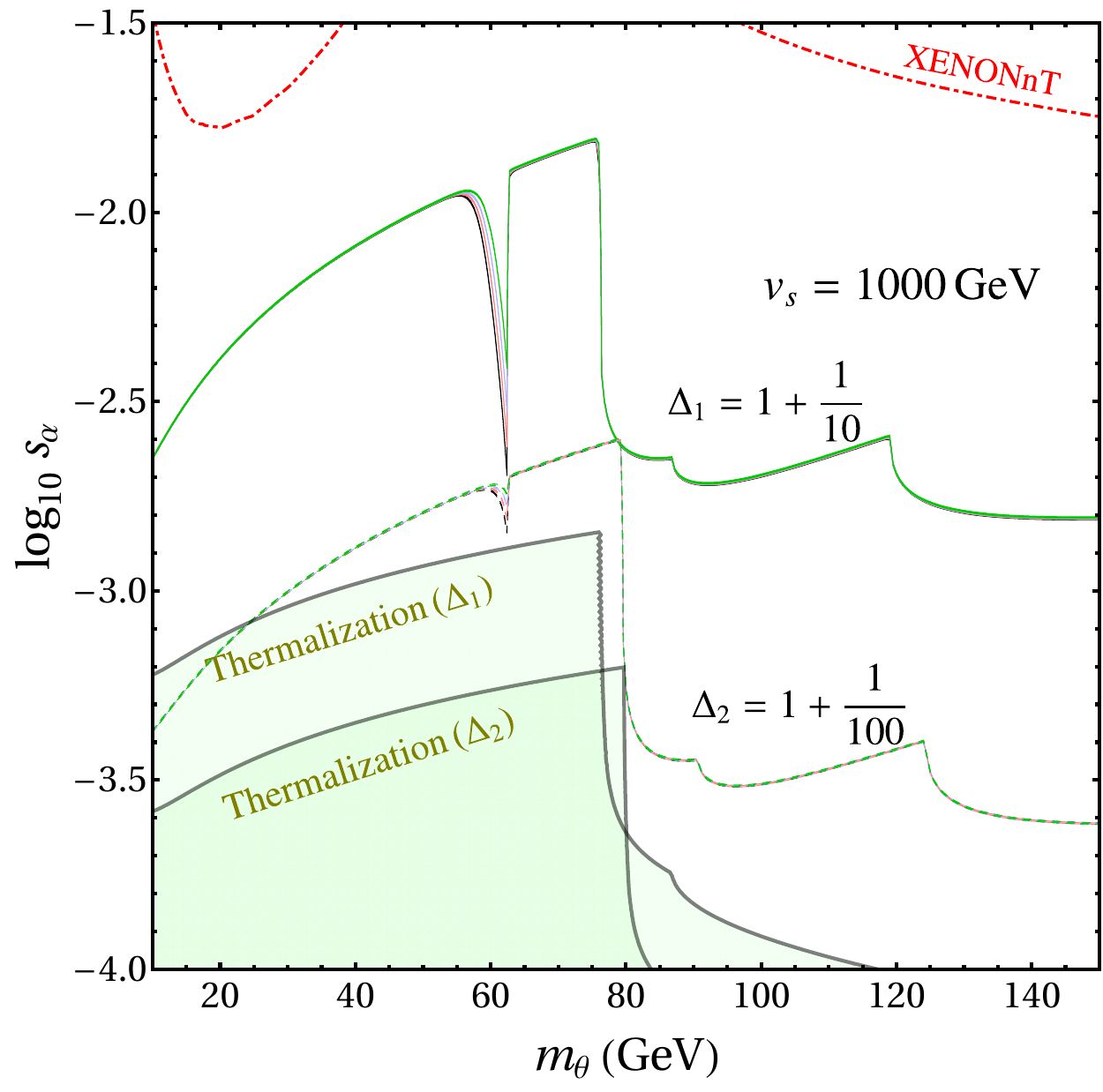}
\par\end{centering}
\caption{\label{fig:Z3Z4res}Curves for the relic abundance
in the $\rho$ resonance for different values of the mass splitting
in Eq.~\eqref{eq:res_condition}, $\Delta_{1}=1+1/10$ (solid), $\Delta_{2}=1+1/100$ (dashed) and $\protect\vs=10^{2},\,10^{3}\,\protect\gev$ (left, right). Experimental constraints from invisible Higgs decays (blue), XENON1T experiment \citep{Aprile:2018dbl} (gray), the projection for XENONnT \citep{Aprile:2015uzo} (red dot-dashed line) and the thermalization condition (green) are shown. Same color code for the \textit{minimal models}
as in Fig.~\ref{fig:Scandeltamixing}.}
\end{figure}

In Fig.~\ref{fig:Z3Z4res} we display the mixing angle versus the
DM mass setting the resonance mass condition in Eq.~\eqref{eq:res_condition}.
We show plots for two different values of the VEV, $\vs=100\,\gev$
in the left panel, $\vs=1000\,\gev$ in the right panel. The case
of $m_{\theta}\simeq m_{h}/2$ corresponds to the DM being in the
resonance of both CP-even scalars, which are almost degenerate. It
can be observed how once $\mdm$ is close to the mass threshold $m_{W},\,m_{Z}$
or $m_{h}$, new channels $\theta\theta\rightarrow WW,\,ZZ,\,hh$
open up. We conclude that the differences in the required mixing angle
among the \textit{minimal models} are not significant in the resonances
of $h$ or $\rho$, so it is very difficult to disentangle them.

\begin{figure}[H]
\begin{centering}
\includegraphics[scale=0.26]{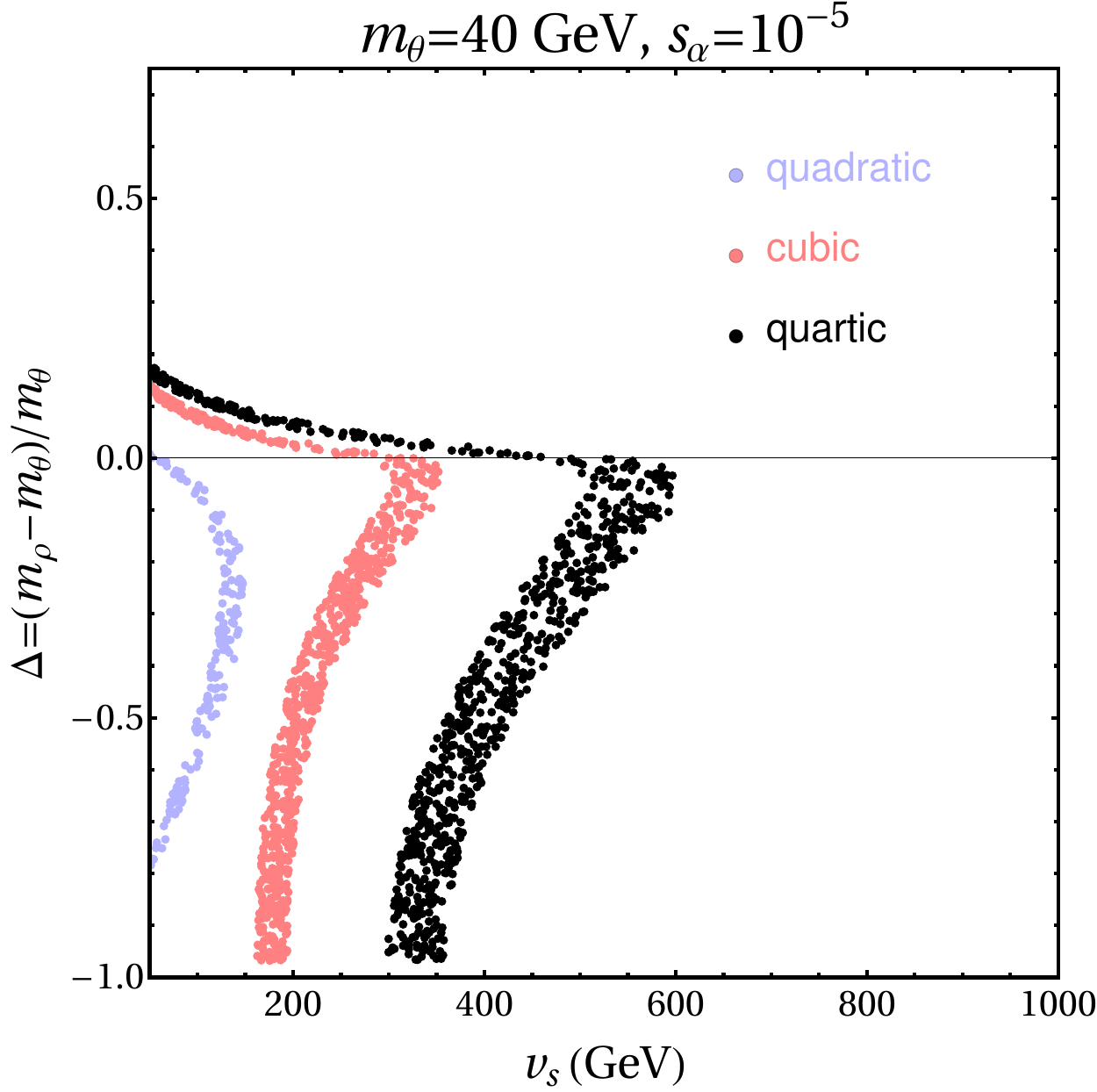}\includegraphics[scale=0.26]{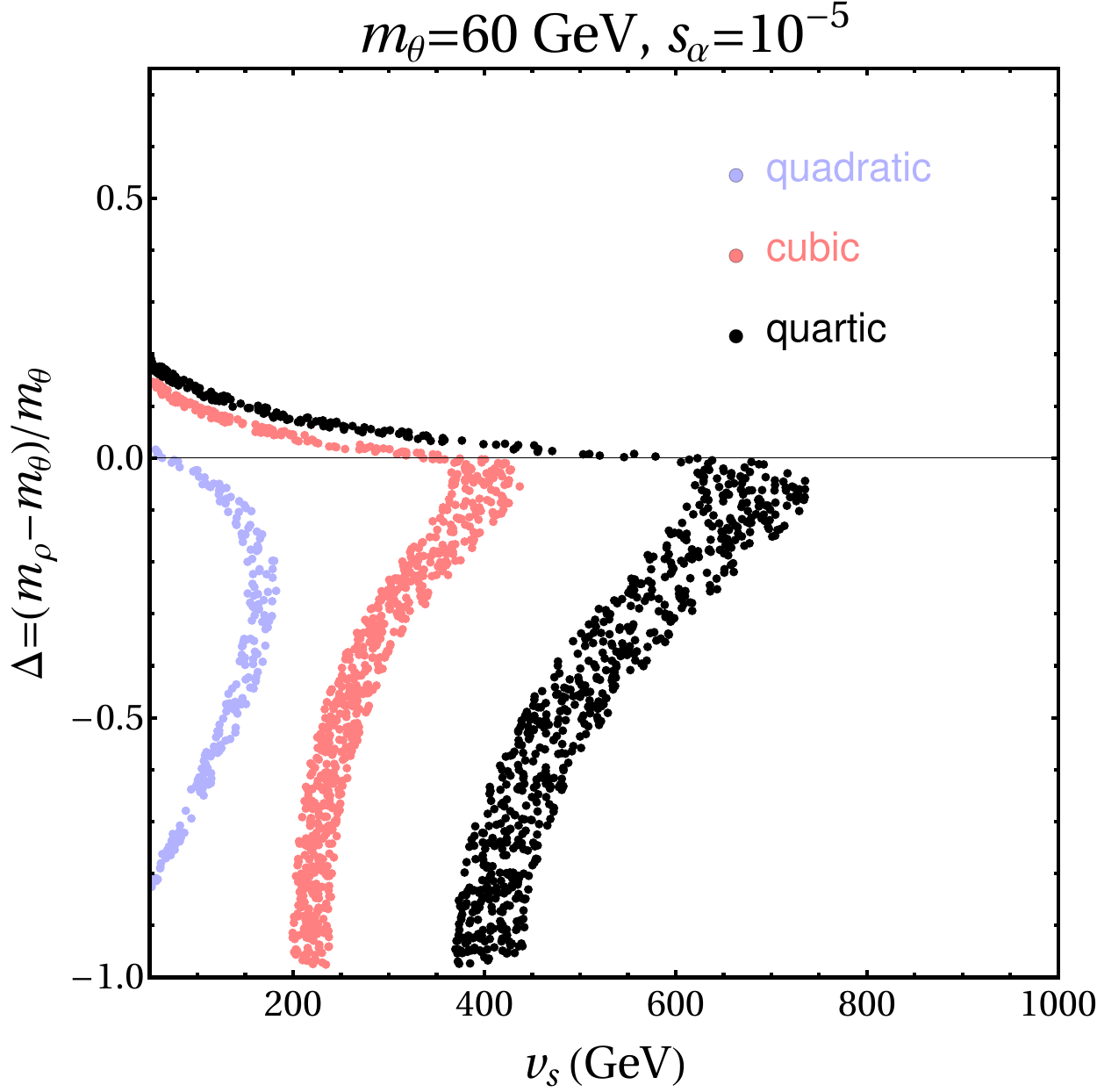}\includegraphics[scale=0.26]{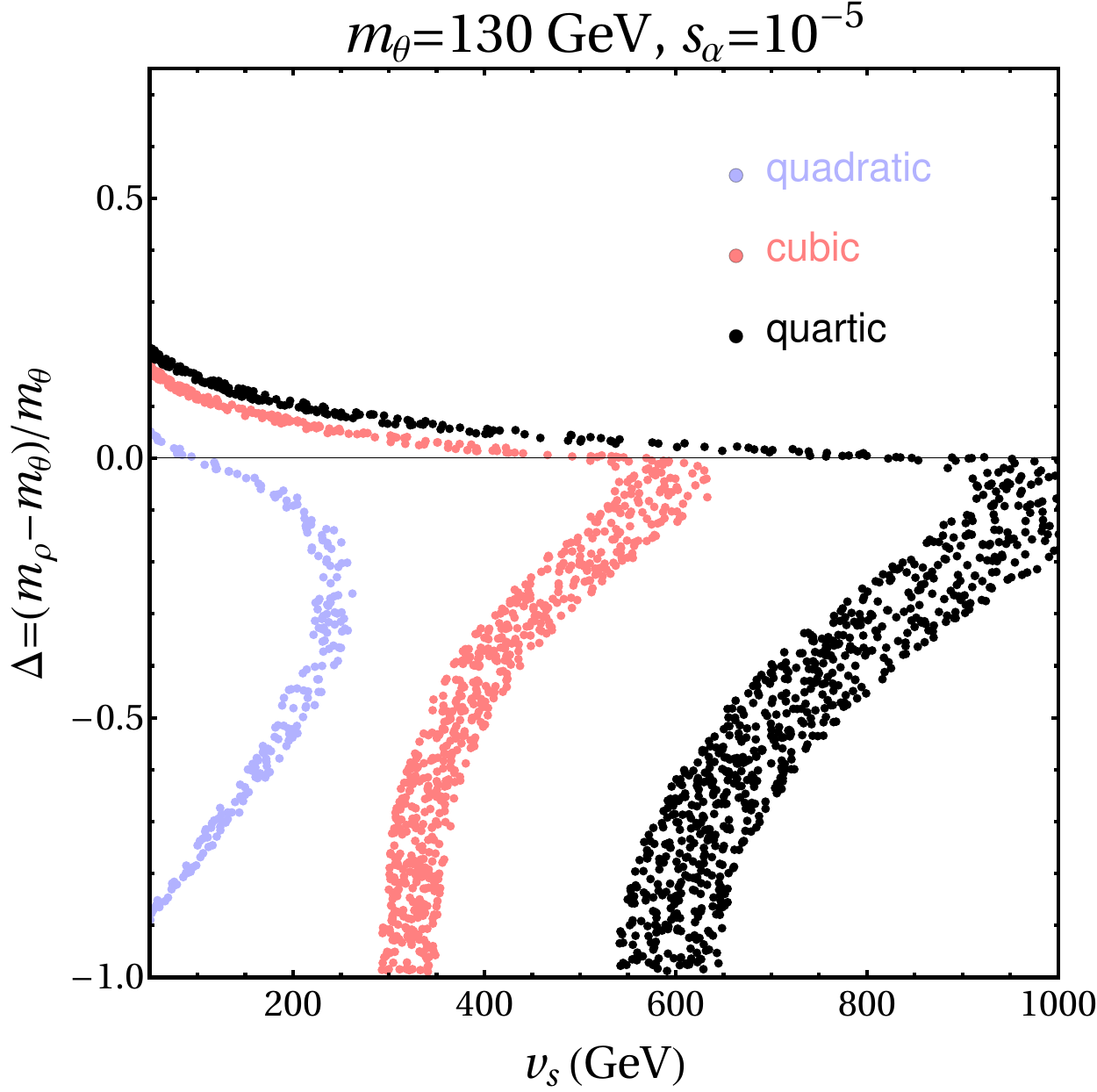}
\par\end{centering}
\caption{\label{fig:delta_vs}Scan in $\Delta$ and $\protect\vs$ for fixed
values of $\protect\mdm$ and $\protect\sa$. The points fulfill the
condition $0.5\leqslant\Omega/\Omega_{\text{obs}}\leqslant1$. We
also impose the XENON1T and the invisible Higgs decay constraints.
The same color code as in Fig.~\ref{fig:Scandeltamixing} is applied.}
\end{figure}

However, the opposite happens in the FDM and SDM scenarios, where
the differences among models are clearly visible. This can also be
observed in Fig.~\ref{fig:delta_vs}, where we plot the normalized
mass splitting versus the VEV $v_{s}$ for the \textit{minimal models}
and different values of the DM mass ($\mdm=40,\,60,\,130\,\gev$ from
left to right) with a fixed (small) mixing angle $s_{\alpha}=10^{-5}$.
Notice that the SDM scenario can also happen for the cubic and quartic
models when the value of $v_{s}$ is increased.

\begin{figure}[H]
\begin{centering}
\includegraphics[scale=0.3]{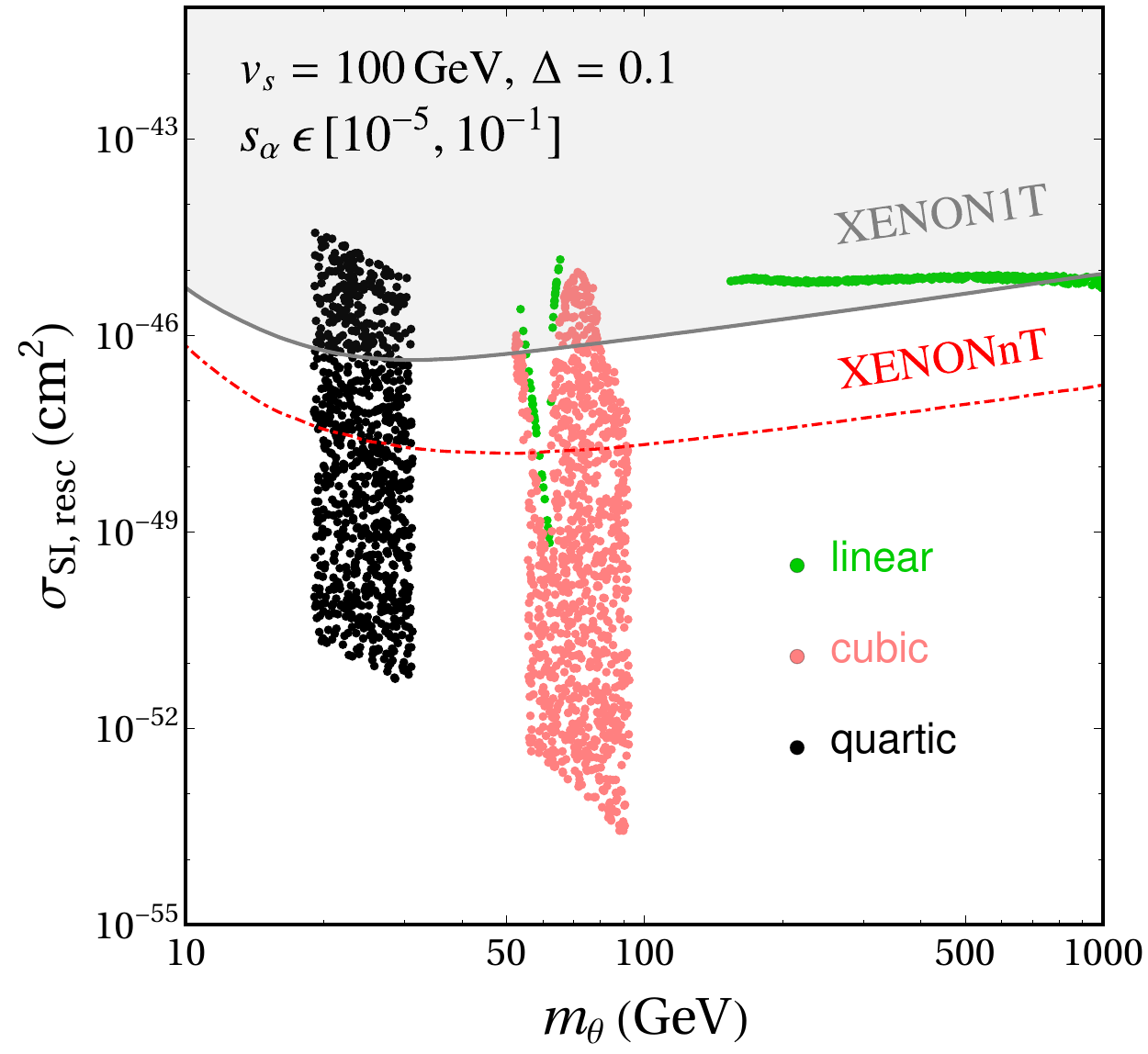}\includegraphics[scale=0.3]{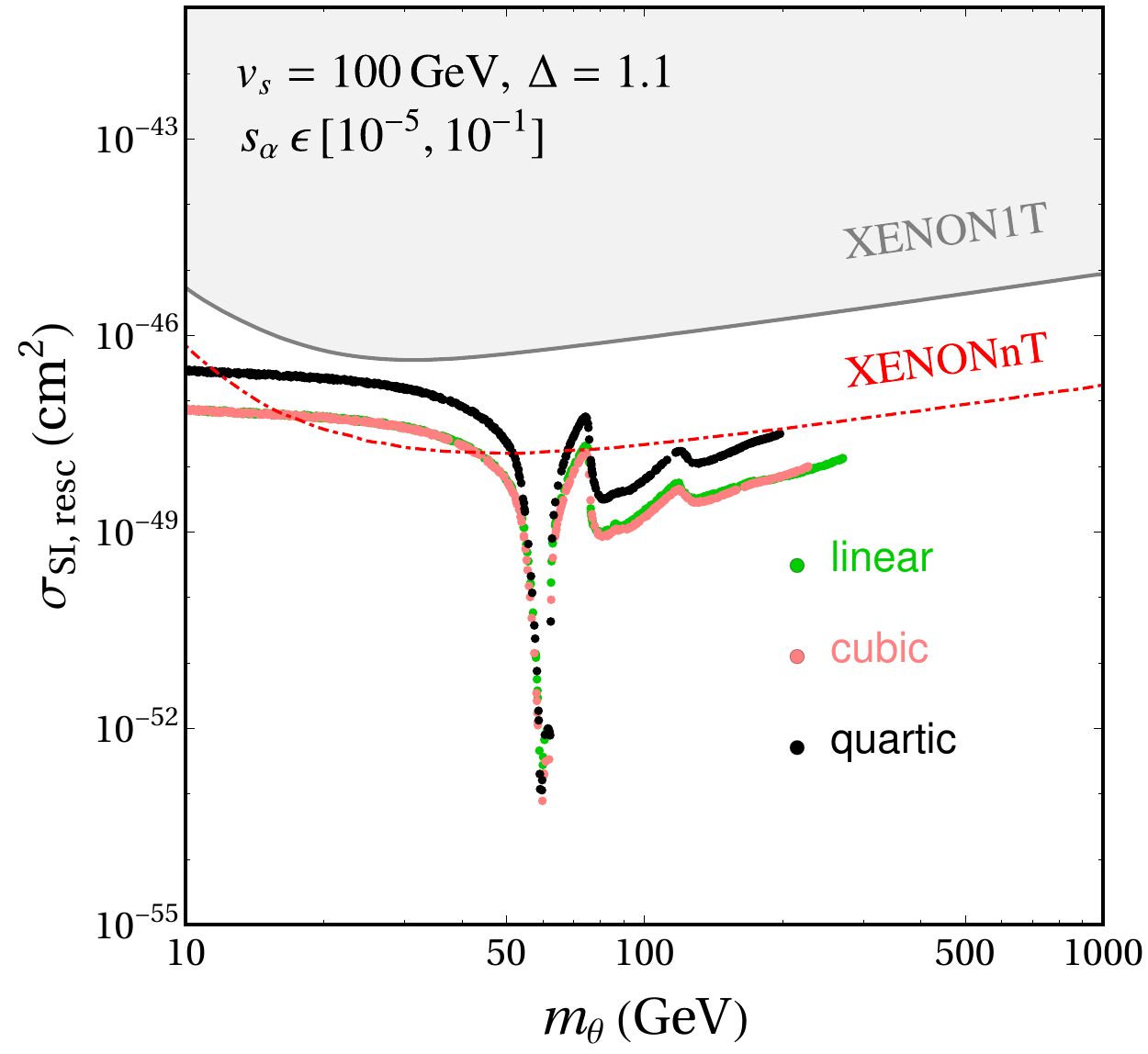}
\par\end{centering}
\caption{\label{fig:DDxsecmthetavs100}Scan in the rescaled spin-independent DD cross section signal
for the \textit{minimal models} versus the DM mass, for mixing in
the range $\protect\sa\,\epsilon\,[10^{-5},\,10^{-1}]$, and fixed
values $\protect\vs=100\,\protect\gev$ and $\Delta=0.1$ (left, FDM)
$\Delta=1.1$ (right, $\rho$ resonance). Constraints from perturbativity
and invisible Higgs decay have been taken into account. We also plot
in gray the exclusion region from the current XENON1T experimental
limit \citep{Aprile:2018dbl} and its projection \citep{Aprile:2015uzo}
as a red dot-dashed line.}
\end{figure}

Now we perform a scan for the different models in the parameters
\[
\sa\,\epsilon\,[10^{-5},\,10^{-1}]\,,\quad\mdm\,\epsilon\,[10,\,1000]\,\gev,
\]
for fixed $\vs=100\,\gev$ and $\Delta$ (FDM, $\Delta=0.1$; $\rho$
resonance $\Delta=1.1$) accepting the points for which $0.5\leqslant\Omega/\Omega_{\text{obs}}\leqslant1$.
In Fig.~\ref{fig:DDxsecmthetavs100} we depict the results: the rescaled
spin-independent DD cross section $\text{\ensuremath{\sigma_{SI,\,\text{resc}}}}=(\Omega/\Omega_{\text{obs}})\sigma_{SI}$
for the different models, along with the current and future limits
of the XENON1T experiment \citep{Aprile:2018dbl,Aprile:2015uzo}.
Notice that for the parameter values considered, in the case of SDM
($\Delta=-0.1$, for instance) the DM is under-abundant, so we do
not show this case. Remember that for the quadratic model in the zero-momentum
limit there is an exact cancellation at tree level in the DD cross
section (see Table \ref{tab:effDMnucleoncoupling}). Therefore, its
dominant contribution is at one loop and is very suppressed. We can
see that in the FDM region $(\Delta=0.1)$, assuming astrophysics
are under control (say, the standard halo model), a precise-enough
positive measurement of a DD signal could allow one to distinguish among
the \textit{minimal models}. This is not the case for the region close
to the $\rho$ resonance $(\Delta=1.1)$, and also close to the
Higgs resonance ($m_{\theta}\simeq60$ GeV), where the required parameters
are quite insensitive to the \textit{minimal model} considered. 

We have checked that in all the \textit{minimal models} there is a significant temperature dependence of the thermally averaged annihilation
cross section in the resonance scenario, allowing to avoid indirect detection (ID) bounds~\citep{Fermi-LAT:2016uux}. This was also found in the case of the quadratic model in Ref.~\citep{Arina:2019tib}. However, in the SDM regime the variation on the temperature is more subtle (see next section). Of course, ID bounds do not constrain FDM, given the absence of DM annihilations at zero temperature in this case.

Finally, we have also studied the DM self-interacting cross section
for each model and found that the maximum values were reached in the
resonance with the scalar $\rho$. However, in the parameter space
considered the values of the cross section were much smaller than
the ones constrained by clusters, $\lesssim1\,{\rm cm^{2}/g}$ \citep{Tulin:2017ara}.

\subsubsection{Light dark matter\label{sec:light-DM}}

In this section we discuss the possibility of light DM (e.g. at the sub-$\gev$ scale) in the \textit{minimal models}\footnote{Ref.~\citep{Hara:2021lrj} also discusses light thermal DM candidates in similar scenarios as those considered in this work.}.  Let us analyze in the following the different freeze-out scenarios described in Sec.~\ref{subsec:Dark-matter-relic}. We impose that the DM candidates are in thermal equilibrium in the Early Universe. Furthermore, constraints from DD experiments, namely CRESST-III and DarkSide-50~\citep{CRESST:2019jnq,DarkSide:2018bpj}, and invisible Higgs decays are applied.

\begin{enumerate}
\item In the $\rho$ resonance, the lightest possible mass for the DM candidate can be read from Fig.~\ref{fig:combinedplot} and it could be at the sub-GeV scale. As discussed in Sec.~\ref{subsec:Comparison-of-minimal} (Fig.~\ref{fig:Z3Z4res}), it is difficult to disentangle the \textit{minimal models} in the resonance condition, therefore the same lower bound for the DM mass applies for the rest of the models. Constraints on the mixing angle coming from meson decays into invisible states ($B\rightarrow K\rho$ and $K\rightarrow\pi\rho$~\citep{Winkler:2018qyg}) forbid lower DM masses. 

\item In the SDM/FDM scenarios, restrictions on the mixing angle come basically from existing limits on light scalars that mix with the Higgs boson (see Fig.~8 in Ref.~\citep{Winkler:2018qyg}). For the SDM regime, ID bounds must also be considered.

\end{enumerate}

\begin{figure}[H]
\begin{centering}
\includegraphics[scale=0.3]{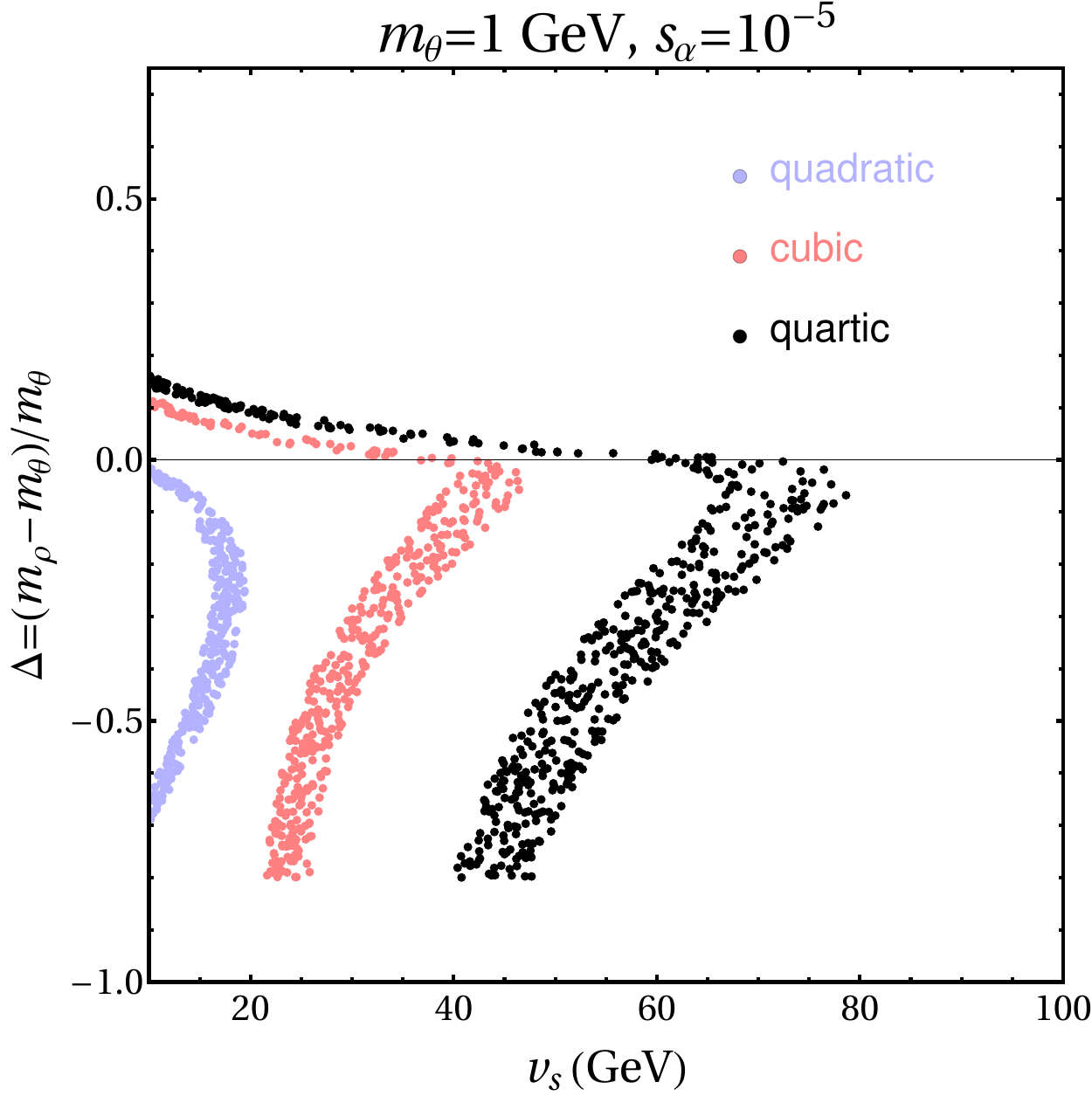}\includegraphics[scale=0.3]{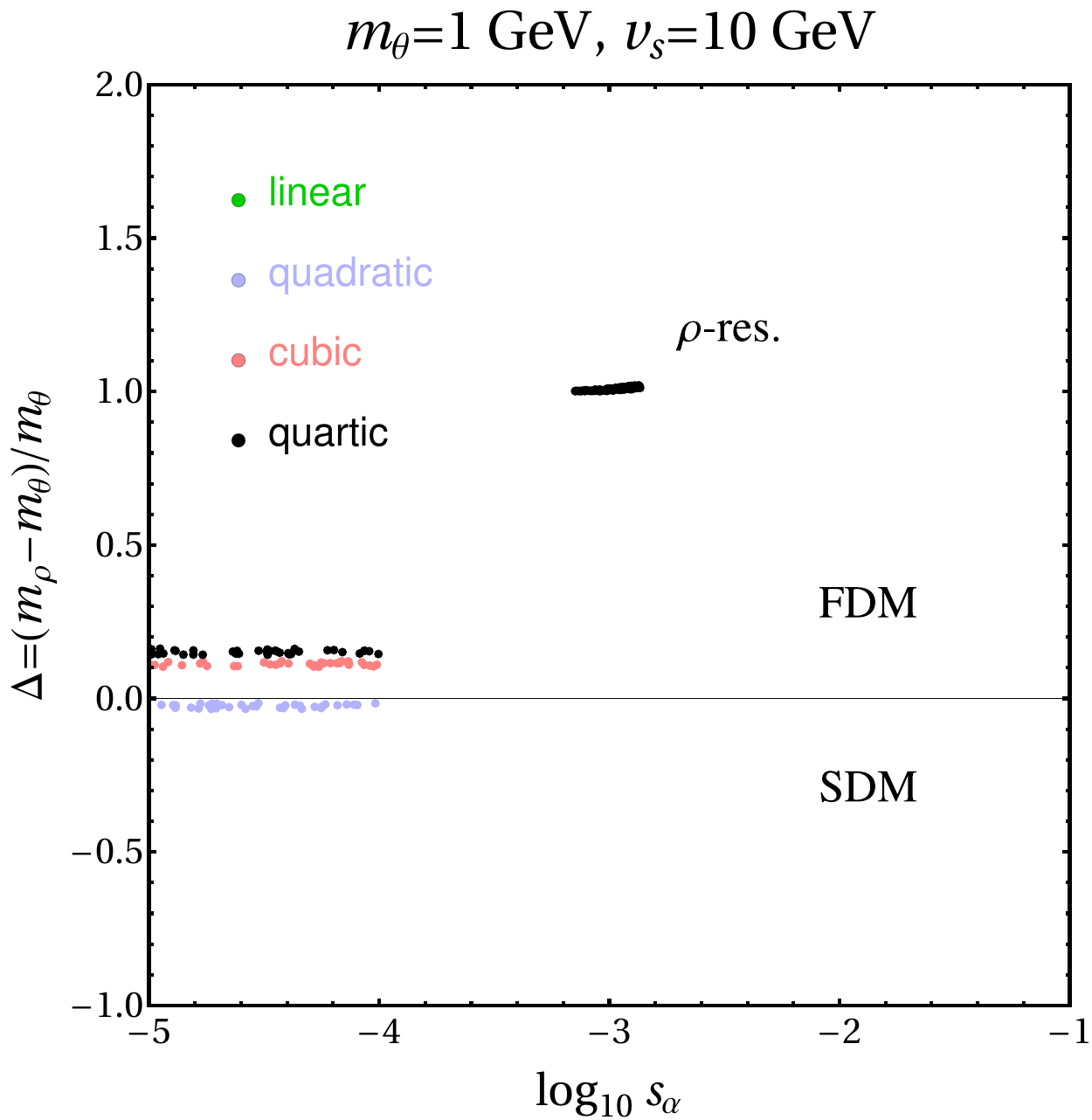}
\par\end{centering}
\caption{\label{fig:lightDMplot}Scan in $\Delta$ and $\vs$ $(\sa)$ with fixed $\mdm=1~\gev$ and $\sa=10^{-5}$ $(\vs=10~\gev)$ for the \textit{minimal models} on the left (right) panel. Constraints from light scalars that mix the Higgs, Fermi-LAT and invisible Higgs decay have been taken into account. All
the points fulfill the relic abundance condition $0.5\leqslant\Omega/\Omega_{\text{obs}}\leqslant1$.}
\end{figure}

As an illustrative example, in Fig.~\ref{fig:lightDMplot} we take $m_{\theta}=1~\gev$ and perform an scan for the \textit{minimal models} where the points fulfill the condition $0.5\leqslant\Omega/\Omega_{\text{obs}}\leqslant1$: on the left panel we plot $\Delta$ versus $\vs$ with $\sa=10^{-5}$; on the right panel we depict $\Delta$ versus $\sa$ with $\vs=10~\gev$. For this value of $\vs$, the constraint from the invisible Higgs decay almost excludes the $\rho$ resonance solution, and ID bound from Fermi-LAT~\citep{Fermi-LAT:2016uux} forbids SDM for $\Delta\lesssim-0.1$.

\section{Beyond \textit{minimal models}\label{sec:beyond-minimal}}

Here we will try to go a bit beyond the \textit{minimal models} discussed
above, which involved just one dominant symmetry breaking term. We
do so by combining pairs of \textit{minimal models}, that is including
simultaneously two symmetry breaking terms. The main goal is to study
if the parameter space opens up. However, notice that the inclusion
of the additional couplings (even if real) can lead to SSB of DCP
and spoil the DM candidate, as it would not longer be stable. Therefore,
we should require that this does not happen, which leads to important
constraints on the parameter space of these non\textit{-minimal models.}

\subsection{About spontaneous dark CP violation \label{subsec:About-Spontaneous-CP}}

In this section we study the case in which the potential contains two explicit $U(1)$ 
symmetry breaking terms, which we write as 
\begin{equation}
V_{\mathrm{xsb}}=\tilde{a}\, S^n + \tilde{b}\, S^{n'}+\mathrm{H.c.}\,,\label{eq:Vbrn1}
\end{equation}
where $n,\,n'=1,\,2,\,3,\,4$ correspond to the linear, quadratic, cubic
and quartic model interactions, and $\tilde{a}$ and $\tilde{b}$ are generically the couplings of 
the two interactions we consider. Without loss of generality we take $n<n'$. 

DCP conservation, defined as $S\rightarrow S^*$, requires that the two couplings $\tilde{a}$ and $\tilde{b}$ are real. To avoid SSB of DCP we should 
also require that the VEV of $S$ is invariant, $\langle S\rangle = \langle S^*\rangle$,
and therefore, $\langle S\rangle$ must be real. Moreover, since the rest of the 
Lagrangian is invariant under phase transformations, we can make a redefinition 
$S\rightarrow -S$ \comment{, which can change the signs of $a$ and/or $b$,} and take 
always $\langle S\rangle$ real and positive. Then, to explore the conditions under which DCP 
is not spontaneously broken it is convenient to use the exponential parametrization 
(see Refs.~\citep{Haber:2012np,Branco:1999fs})
\begin{equation}
S=\frac{1}{\sqrt{2}}\left(\vs+\sigma'\right)e^{iG/\vs}\,,\label{eq:exponential-parametrization}
\end{equation}
because in this parametrization the field $G$ only appears in the symmetry breaking part of the 
Lagrangian, Eq.~(\ref{eq:Vbrn1}). Then, according to the discussion above, we require that the 
global minimum of the potential is found at $\langle \sigma'\rangle=0$ (this will fix the value of $\vs$ 
in terms of all the parameters of the potential) and 
$\langle G\rangle=0$ so that $ \sqrt{2}\langle S \rangle = \vs >0$. In this parametrization
the potential  contains the terms
\begin{equation}
V_{\mathrm{xsb}} \supset V_{\mathrm{sb}} \equiv 
\vs^{4}\,\left(a\,\cos\left(n\frac{G}{v_{s}}\right)+b\,\cos\left(n^{\prime}\frac{G}{v_{s}}\right)\right)\,
,\label{eq:Vbrn}
\end{equation}
where we have defined the dimensionless couplings 
\begin{equation}
a\equiv \frac{2}{\vs^4}\left(\frac{\vs}{\sqrt{2}}\right)^n \tilde{a}\,\, , \qquad
b\equiv \frac{2}{\vs^4}\left(\frac{\vs}{\sqrt{2}}\right)^{n'} \tilde{b}\, , \label{eq:Vbrnab} 
\end{equation}
which are expressed in terms of the original couplings of the Lagrangian as given in Table~\ref{tab:tabl_CP-1}.

\comment{The results apply also to the cases in which the full symmetry breaking
potentials, Eqs.~\eqref{eq:complex-scalar-potential-V1} and \eqref{eq:complex-scalar-potential-V2}
(instead of Eqs.~\eqref{eq:complex-scalar-potential-V1-1} and \eqref{eq:complex-scalar-potential-V2-1})
are realized, with the substitutions of \footnote{Other symmetry breaking terms coming from higher dimensional operators
proportional to different powers of $S$ can also be trivially included
in the analysis. }
\begin{equation}
\mu^{3}\rightarrow\mu^{3}+\frac{1}{2}\mu_{H1}v^{2}+\frac{1}{2}\mu_{1}v_{s}^{2}\,,
\end{equation}

\begin{equation}
\mu_{S}^{2}\rightarrow\mu_{S}^{2}+\frac{1}{2}\lambda_{H2}v^{2}+\frac{1}{2}\lambda_{2}v_{s}^{2}\,.
\end{equation}
}
\begin{table}[H]
\begin{centering}
\begin{tabular}{|c|c|c|}
\hline 
 $n,\,n^\prime$ & \textit{Minimal model} & $a,b$\tabularnewline
\hline
\hline 
$1$ & Linear & $\mu^{3}/\left(\sqrt{2}\,v_{s}^{3}\right)$\tabularnewline
\hline 
$2$ & Quadratic & $\mu_{s}^{2}/\left(2\,v_{s}^{2}\right)$\tabularnewline
\hline 
$3$ & Cubic & $\mu_{3}/\left(2\sqrt{2}\,v_{s}\right)$\tabularnewline
\hline 
$4$ & Quartic & $\lambda_{4}/4$\tabularnewline
\hline 
\end{tabular}
\par\end{centering}
\caption{\label{tab:tabl_CP-1}Relation of the effective parameters $a$ and $b$ defined in 
Eqs.~(\ref{eq:Vbrn1},\ref{eq:Vbrn}--\ref{eq:Vbrnab}) and
the explicit symmetry breaking couplings in the potential for the \textit{minimal models}.}
\end{table}

\comment{It is easy to see that if $a>0$, $G=0$ cannot be the global minimum of the potential. In fact
$V_\mathrm{sb}(G=0)=v_{s}^4(a+b)$, while $V_\mathrm{sb}(G=2\pi\vs/n')=v_{s}^4(a \cos(2\pi n/n')+b)$, which is smaller 
than $v_{s}^4(a+b)$ if $a>0$. 
If $a\leq 0$ one must first check that $G=0$ is a local minimum and, then, that it is the global minimum 
by comparing with 
the value of $V_\mathrm{sb}(G)$ at the other minima. We thus find that the conditions for DCP conservation are}

Then, to avoid spontaneous violation of DCP one has to require that $G=0$ is the global minimum of 
$V_\mathrm{sb}$ in Eq.~\eqref{eq:Vbrn}. Obviously, the first derivative of $V_\mathrm{sb}$ in $G=0$ is zero, 
while the second derivative is $-\vs^2 (n^2 a+n^{'2} b)$. Therefore, to have a local minimum at $G=0$ one should 
require $b \le -a(n^{2}/n'^{2})$. Then, one has to check that it is the global minimum of $V_\mathrm{sb}(G)$
by comparing it with other minima, which has to be done case by case. We thus find that the conditions for 
DCP conservation are
\begin{equation}
a\leq0\quad\mathrm{and}\quad\begin{cases}
b\leq0 & (n,n')=(2,3),(3,4)\\
b\le-a(n^{2}/n'^{2}) & (n,n')\not=(2,3),(3,4)
\end{cases}
\end{equation}

\comment{\begin{table*}
\begin{centering}
\caption{\label{tab:tabl_CP}Upper limits on $b$ such that the dark symmetry
DCP is preserved by the vacuum. It is always needed that $a\leqslant0$.}
\begin{tabular}{ccccc}
\hline\noalign{\smallskip}
& \textbf{$n'=1$} & \textbf{$n'=2$} & \textbf{$n'=3$} & \textbf{$n'=4$}  \\
\noalign{\smallskip}\hline\noalign{\smallskip}
\textbf{$n=1$} & No restriction & $b\leqslant-a/2^{2}$ & $b\leqslant-a/3^{2}$ & $b\leqslant-a/4^{2}$ \\
\textbf{$n=2$} & - & No restriction & $b\leqslant0$ & $b\leqslant-a(2/4)^{2}$ \\
\textbf{$n=3$} & - & - & No restriction & $b\leqslant0$ \\
\textbf{$n=4$} & - & - & - & No restriction \\
\noalign{\smallskip}\hline
\end{tabular}
\par\end{centering}
\end{table*}
}

\begin{figure}[H]
\begin{centering}
\includegraphics[scale=0.35]{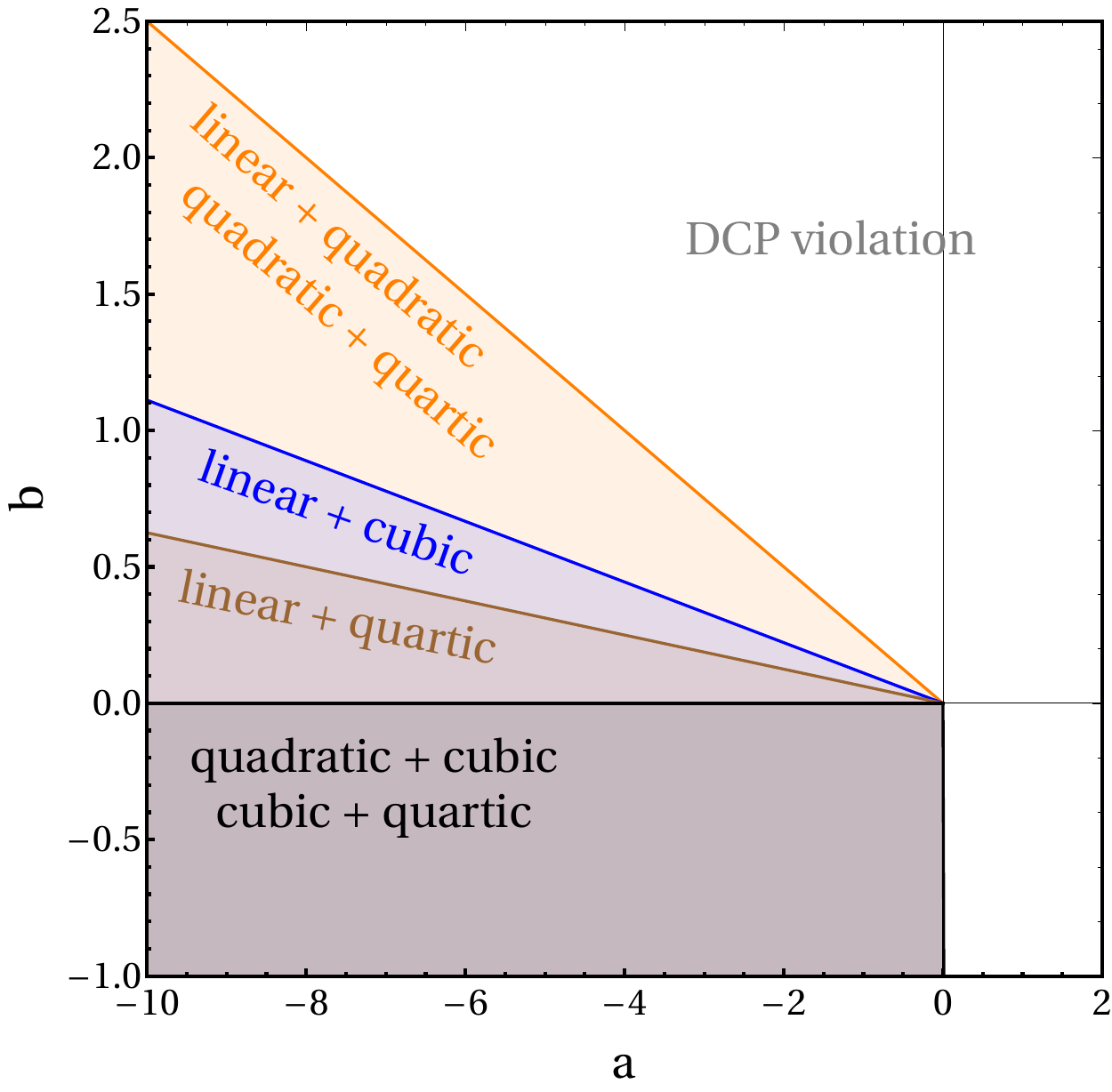}
\par\end{centering}
\caption{\label{fig:DCP}Upper limits on the effective symmetry breaking terms
$a$ and $b$ for DCP conservation, as described in Eq.~\eqref{eq:Vbrn}
for all the possible combinations of two \textit{minimal models}. Combinations of models which appear with 
the same color share the same DCP conserving parameter region.}
\end{figure}

These constraints are displayed in Fig.~\ref{fig:DCP} in the $\left(a,b\right)$
plane and, when expressed in terms of the explicit symmetry breaking
parameters of the general potential in 
Eqs.~(\ref{eq:complex-scalar-potential}--\ref{eq:complex-scalar-potential-V4}),
must be added to the theoretical constraints mentioned in Sec.~\ref{subsec:minimal-models}. 
\comment{This argument severely restricts the parameter space of models with two symmetry breaking terms.} 

\begin{figure}[H]
\begin{centering}
\includegraphics[scale=0.3]{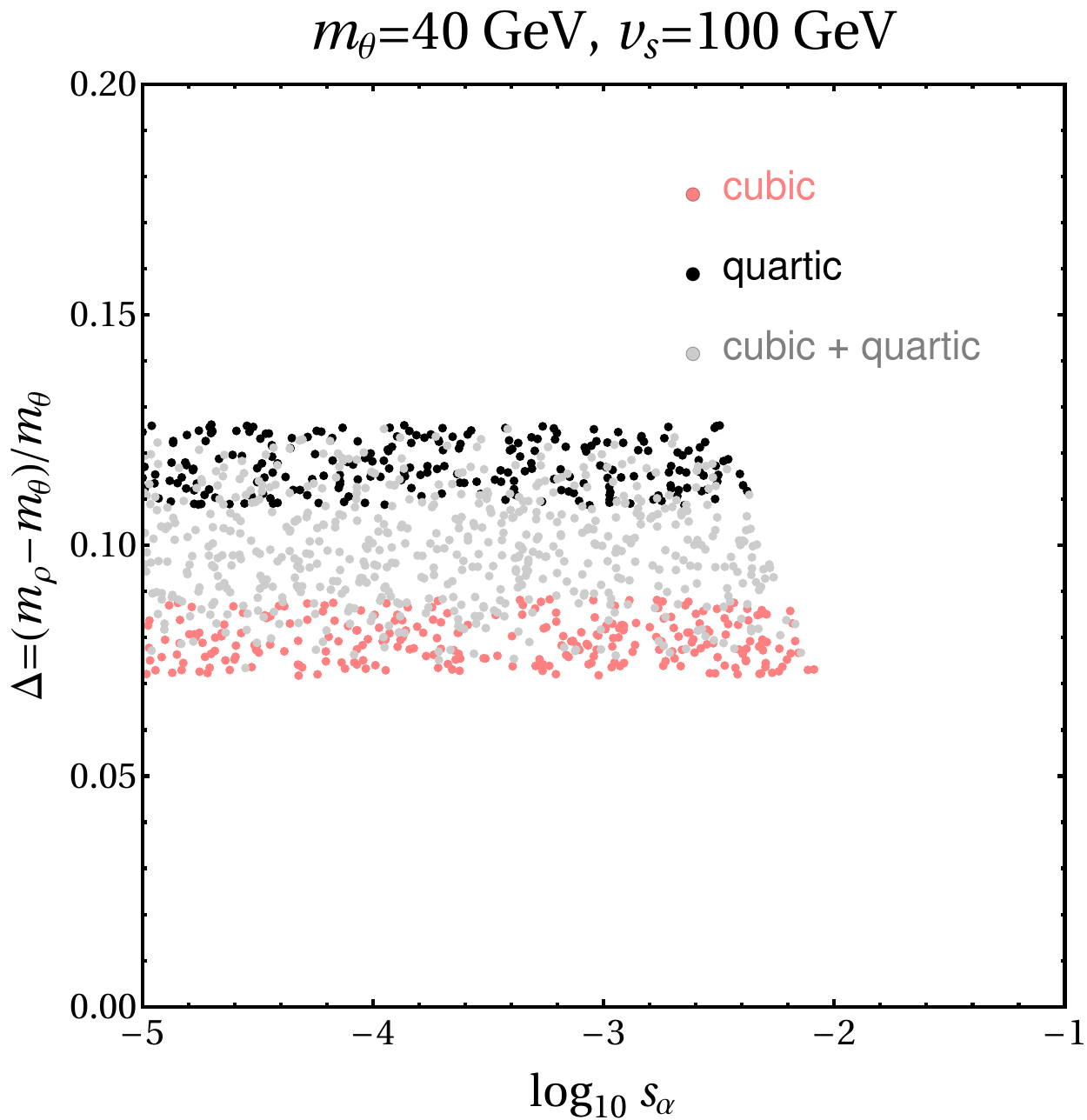}~\includegraphics[scale=0.31]{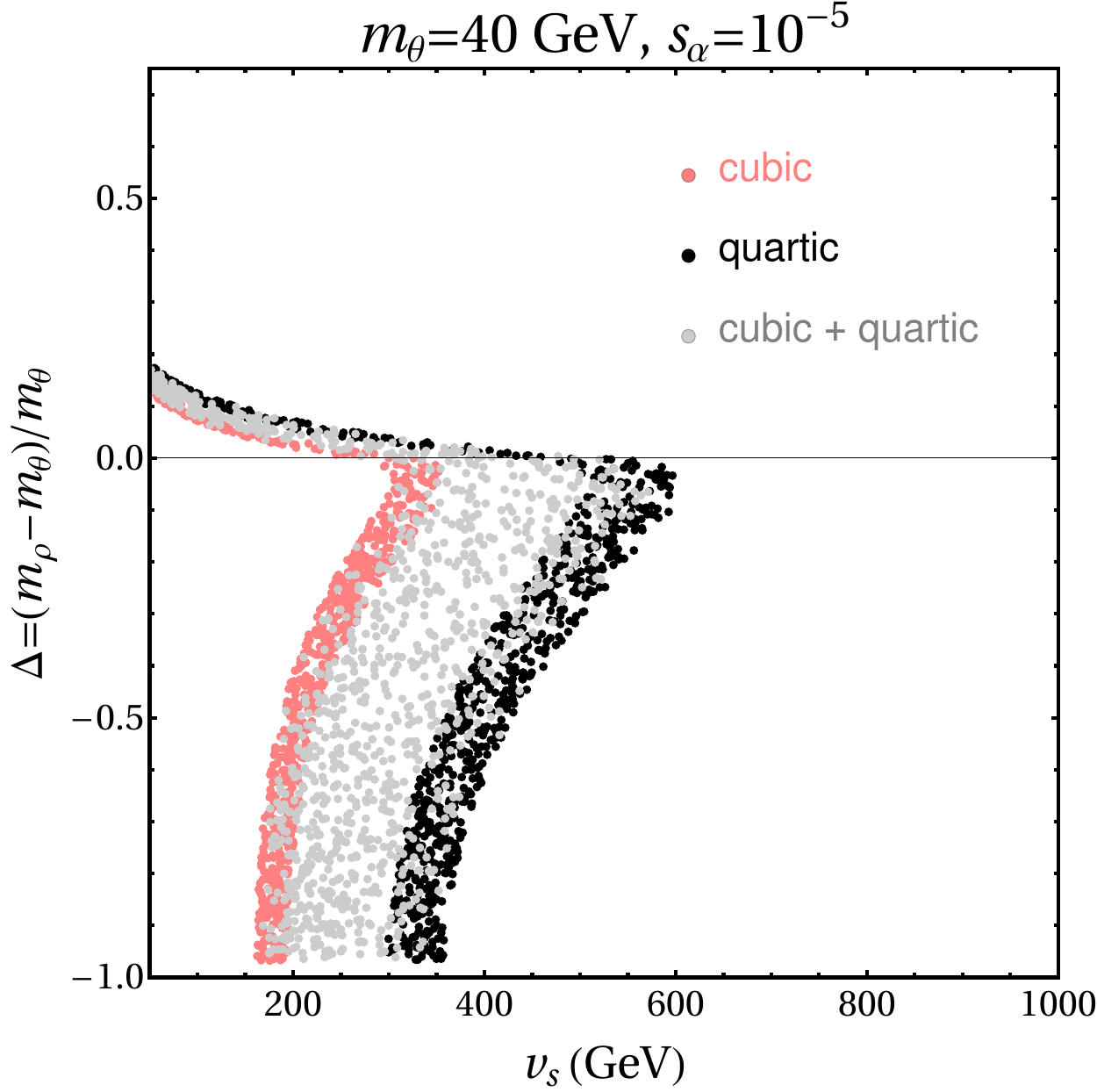}
\par\end{centering}
\caption{\label{fig:2breaking34}Left) Scan in $\Delta$ and $\protect\sa$
for the cubic and quartic models around the region of $\Delta\gtrsim0$
(FDM), similar to Fig.~\ref{fig:Scandeltamixing} (left). Points
in gray correspond to the region of parameter space of the model which
is the combination of both interactions and preserves DCP. Right)
Scan in $\Delta$ and $v_{s}$, similar to Fig.~\ref{fig:delta_vs}
(left), enlarged with the combined model.}
\end{figure}

As an illustrative example of the consequences of combining two \textit{minimal
models}, in Fig.~\ref{fig:2breaking34} we plot the results in the
planes $\left(\sa,\Delta\right)$ on the left and $\left(\vs,\Delta\right)$
on the right for having a suitable DM candidate when considering the
cubic and quartic symmetry breaking terms, both alone and together.
The results show that including both terms allows values of $m_{\rho}$
and $v_{s}$ in between the individual cases (\textit{minimal models});
that is the region covered by the gray dots in the figure. Analogous
results were found when considering the other combinations of \textit{minimal
models}. 

In conclusion, having two symmetry breaking terms enlarge the parameter
space for a good DM candidate to regions bounded by the \textit{minimal
models}. This is a consequence of the additional constraints required to avoid
the spontaneous violation of DCP shown in Fig.~\ref{fig:DCP}.

\section{The pseudo-Nambu-Goldstone Boson limit: an EFT approach \label{sec:pGB-limit}}

\subsection{Effective operators}

For the following discussion it is useful to consider the exponential
parametrization, Eq.~\eqref{eq:exponential-parametrization}, in
terms of the radial mode $\sigma'$ and the angular mode $G$. Let
us now assume that $S$ takes a large VEV as compared to the explicit
symmetry breaking terms. Then, the mass of the angular mode $G$ is
much smaller that the symmetry breaking scale $m_{G}\ll v_{s}$ and
it can be considered a PNGB. At low energies, the scenario which involves
a PNGB $G$ stabilized by a discrete symmetry (DCP) $G\rightarrow-G$
can be parametrized by 
\begin{equation}
\Delta\mathcal{L}=\frac{1}{2}(\partial G)^{2}+\frac{c_{{\rm G}}}{v_{s}^{2}}\left(|H|^{2}-\frac{v^{2}}{2}\right)(\partial G)^{2}-\frac{1}{2}m_{{\rm G}}^{2}G^{2}+\lambda_{{\rm G}}G^{4}+\lambda_{{\rm HG}}\left(|H|^{2}-\frac{v^{2}}{2}\right)G^{2}\,,\label{eq:EFTDM}
\end{equation}
where the PNGB mass $m_{{\rm G}}$, the quartic coupling $\lambda_{{\rm G}}$
and the Higgs portal $\lambda_{{\rm HG}}$ break the shift-symmetry.
These correspond at high energies to linear, quadratic, cubic and
quartic terms in the complex singlet scalar $S$ in the linear parametrization.
Notice that they are not suppressed by derivatives, e.g., they can
give significant contributions for instance in DD. 

Let us parametrize the breaking with the charge $n$ of the field
$S$ (e.g., $V\propto S^{n}+{\rm H.c.}$). We can now proceed to integrate-out
the radial mode $\sigma'$. In terms of the UV parameters, the Wilson
coefficients read
\begin{align}
c_{{\rm G}}= & -s_{\alpha}\frac{v_{s}}{v}\,,\label{eq:c_G}\\
\lambda_{{\rm HG}}= & -c_{{\rm G}}\,\frac{n}{2}\,\frac{m_{{\rm G}}^{2}}{v_{s}^{2}}\,,\label{eq:lambda_HG}\\
\lambda_{{\rm G}}= & \frac{n^{2}}{8}\,\frac{m_{{\rm G}}^{4}}{m_{\sigma'}^{2}v_{s}^{2}}\,,
\end{align}
and $m_{{\rm G}}=m_{\theta}$ is given in Eq.~\eqref{eq:mtheta_general}.
Notice that $m_{{\rm G}}$ and $\lambda_{{\rm G}}$ (and $c_{G}$
and $\lambda_{{\rm HG}}$) are related with each other. We have expressed
these coefficients in terms of physical parameters using also that,
in the limit $\vs\gg v$,
\begin{equation}
s_{\alpha}\simeq\frac{\lambda_{HS}}{2\,\lambda_{S}}\frac{v}{\vs}\,.
\end{equation}
If one considers a certain shift symmetry $G/\text{\ensuremath{\vs}}\rightarrow G/\vs+2\pi/n$
at low energies, the non-derivative terms may be dropped and the allowed
terms in the potential take the form

\begin{equation}
\Delta V=\sum_{n=1}^{4}V_{n}+\sum_{n=1}^{2}U_{n}\,,
\end{equation}
where

\begin{equation}
V_{n}=d_{n}\cos\left(n\frac{G}{v_{s}}\right)\,\qquad\text{and}\qquad U_{n}=c_{n}\left(|H|^{2}-\frac{v^{2}}{2}\right)\left(1-\cos\left(n\frac{G}{v_{s}}\right)\right)\,,
\end{equation}
and we have assumed a renormalizable UV completion, which sets the
upper limit in the sums. These $d_{n}$ (with $n=1,\ldots,4$) low-energy
terms have mass-dimension 4 and correspond at high energies, respectively,
to terms linear, quadratic, cubic and quartic in the complex singlet
scalar $S$ in the exponential parametrization, e.g., respecting respectively
the discrete symmetry DCP, $Z_{2}$, $Z_{3}$, $Z_{4}$. Similarly,
the $c_{n}$ (with $n=1,2$) have mass dimension two and correspond
at high energies, respectively, to terms linear and quadratic with
the Higgs doublet, e.g., respecting respectively the discrete symmetry
DCP and $Z_{2}$. Once expanded, they yield the mass and quartic interactions
of the PNGB.

It is interesting to note that if we integrate by parts the derivative
interaction in Eq.~\eqref{eq:EFTDM} we obtain
\begin{align}
\left(|H|^{2}-\frac{v^{2}}{2}\right)(\partial G)^{2} & \rightarrow-\left(|H|^{2}-\frac{v^{2}}{2}\right)G\partial^{2}G-G\partial_{\mu}\left(|H|^{2}\right)\partial^{\mu}G\nonumber \\
 & \rightarrow m_{G}^{2}\left(|H|^{2}-\frac{v^{2}}{2}\right)G^{2}-G\partial_{\mu}\left(|H|^{2}\right)\partial^{\mu}G\,,\label{eq:partial-integration}
\end{align}
where in the last step we use the Klein-Gordon equation of motion
for $G$, which is correct for $G$ on-shell, as in the case of DD
experiments. If we substitute back in Eq.~\eqref{eq:EFTDM} we find
an additional contribution to the Higgs portal coupling,
\begin{equation}
\lambda_{\mathrm{HG}}\rightarrow\lambda_{\mathrm{HG}}+m_{G}^{2}\frac{c_{{\rm G}}}{v_{s}^{2}}=c_{G}\frac{m_{G}^{2}}{v_{s}^{2}}\left(-\frac{n}{2}+1\right)\label{eq:lambdaHG-shift}
\end{equation}
which exactly cancels for $n=2$ and produces a suppression in DD
experiments, since the other terms always involve the momenta of the
PNGB. This cancellation was already noticed in Ref.~\citep{Alanne:2020jwx}.

Regarding the relic abundance, there are two options depending on
the scale of $v_{s}$:
\begin{itemize}
\item If $v_{s}\ll v$, unless very small couplings are involved, we expect
$m_{h}\gg m_{\sigma'},m_{G}$. In order to explain the relic abundance,
the DM $G$ needs to be at the radial mode ($\sigma'$) resonance,
e.g., $m_{G}\simeq m_{\sigma'}/2\ll v$, or almost degenerate with
the $\sigma',$ e.g. $m_{G}\simeq m_{\sigma'}\ll v$ (for SDM/FDM).
In either case, the small mass splitting between $\sigma'$ and $G$
is not naturally achieved, as the symmetry breaking terms need to
be comparable to $v_{s}$, and therefore, strictly speaking, in that
case $G$ cannot be considered a natural PNGB.
\item If $v_{s}\gg v$, unless very small couplings are involved, we expect
$m_{\sigma'}\gg m_{h},m_{G}$. We can integrate-out the radial mode
$\sigma'$, and, in this case, the prediction is that the PNGB needs
to be at the Higgs resonance to explain the relic abundance, e.g.,
$m_{G}\simeq m_{h}/2$, see Refs.~\citep{Arina:2019tib,Gross:2017dan},
or almost degenerate with the $\sigma',$ e.g. $m_{G}\simeq m_{\sigma'}\ll v$
(for SDM/FDM). In this scenario, the small mass splitting between
$h$ and $G$ is achieved for breakings of the order of the electroweak
scale (which is below $v_{s}$). We can use an effective Lagrangian
with just the PNGB and the Higgs field (see also Refs.~\citep{Balkin:2018tma,Ruhdorfer:2019utl}),
and compare to the results of the complete model (for instance the
relic density and direct detection). 
\end{itemize}

\subsection{Numerical results}

\begin{figure}[H]
\begin{centering}
\includegraphics[scale=0.42]{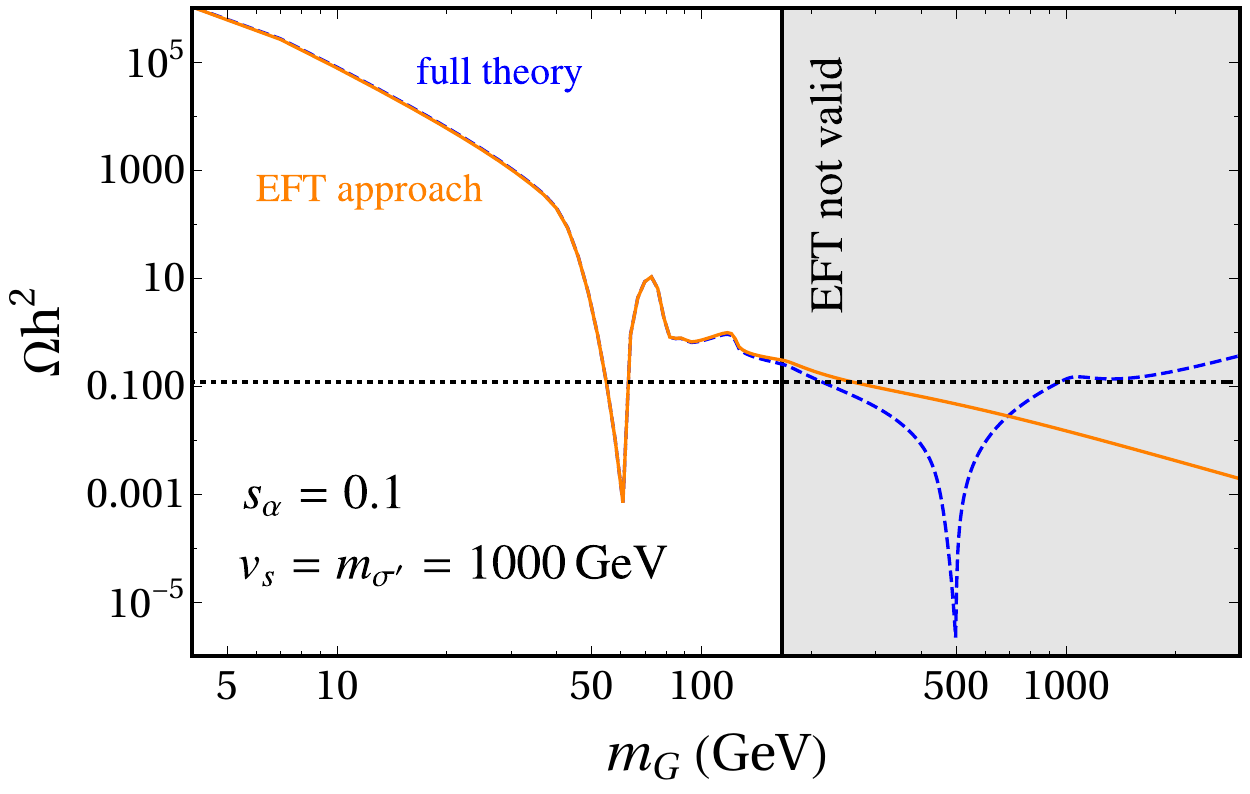}
\par\end{centering}
\caption{\label{fig:Leff-1}Comparison of the relic abundance between the PNGB
effective Lagrangian with the Higgs field, Eq.~\eqref{eq:EFTDM}
(orange line), and the full theory for the quadratic model in Eq.
\eqref{eq:complex-scalar-potential-V2-1}, which also includes the
radial mode $\sigma'$ (blue dashed line). The correct value for the
relic abundance is shown as a black dotted line, e.g., the region
above is excluded and in the region below the PNGB is under-abundant.}
\end{figure}

In Fig.~\ref{fig:Leff-1} we plot the relic abundance versus the
DM mass for both the effective Lagrangian of the PNGB in Eq.~\eqref{eq:EFTDM}
and the full quadratic model which includes the radial mode, with
the symmetry breaking term in Eq.~\eqref{eq:complex-scalar-potential-V2-1}
that gives the mass to the DM candidate.

In the complete model (blue dashed line) we have set $\sa=0.1$ and
$\vs=m_{\sigma'}=10^{3}\,\gev$, where $m_{\sigma'}$ is the mass
of the radial mode. In the effective Lagrangian approach (orange line),
we use the Wilson coefficients $c_{\textrm{G}}$ and $\lambda_{\textrm{HG}}$
from Eqs.~\eqref{eq:c_G} and \eqref{eq:lambda_HG} with the values
for $\sa$ and $\vs$ mentioned before. Notice that, as we are comparing
the effective Lagrangian with the full quadratic model, $n=2$ in
the $\lambda_{\textrm{HG}}$ coefficient. We can see both the Higgs
and the radial resonances in the case of the complete model. The EFT
reproduces very well the full model, including the Higgs resonance,
up to DM masses below $\simeq v_{s}/6$, above which the EFT cannot
be trusted. In the full model, we also see the solution at $m_{G}\simeq m_{\sigma'}\simeq1\,\tev$
(forbidden/secluded, annihilations into $\sigma'$ bosons), which
of course cannot be captured by the EFT. In the full model and in
the EFT we see a small kink at $m_{G}\simeq m_{h}$ corresponding
to the opening of the annihilation channel $GG\rightarrow hh$. A
small kink at $m_{G}=1$ TeV, corresponding to the opening of the
channel $GG\rightarrow\sigma'\sigma'$, can also be seen in the full
model.

\section{Conclusions \label{sec:conc}}

A real scalar singlet, stabilized by a discrete symmetry, is one of
the simplest candidates for DM. However, by combining relic abundance
constraints with direct detection null-results, the allowed parameter
space has been almost completely ruled-out, leaving only a few
allowed spots: at the Higgs boson resonance, or at high masses, where
annihilations into Higgs boson pairs opens up.

In this work we have extended this minimal scenario in the simplest
possible way: a complex scalar singlet, charged under a global $U(1)$
symmetry. If the symmetry is not broken, neither spontaneously nor
explicitly, the allowed parameter space is very similar to that of
the real scalar singlet\footnote{There can be cases with two component DM, which will be studied elsewhere.}.
If the symmetry is preserved at the Lagrangian level but not respected
by the vacuum of the theory there will be an exact Goldstone boson
in the spectrum and, since it is massless, it is dark radiation and
cannot constitute the DM of the Universe. Therefore, the symmetry
must also be broken explicitly at the Lagrangian level, yielding a
pseudo-Nambu-Goldstone Boson as the DM candidate. The explicit symmetry
breaking of the $U(1)$ needs to preserve at least the dark CP that
stabilizes the DM particle. This is the case we have studied in this
work.

We have considered 4 cases with only one explicit breaking term, which
we call linear, quadratic, cubic and quartic. The choices are motivated
by either being the softest possible symmetry breaking term, or by
a discrete symmetry. All these models preserve a dark CP symmetry,
while the quadratic, cubic and quartic models are also invariant under
$Z_{2}$, $Z_{3}$ and $Z_{4}$, respectively. These are the simplest
models in the sense that the breaking involves only the new complex
scalar singlet and it is the softest within each class. Furthermore,
they are quite stable under radiative corrections, as only the case
of the quadratic model generates, at one loop, further symmetry breaking
terms, which are, however, suppressed.

The models are necessarily very predictive, with just four new parameters:
$m_{\theta},m_{\rho},s_{\alpha}$ and $v_{s}$. However, the allowed
parameter space to obtain the correct relic abundance is enlarged
significantly with respect the case of only one real scalar because
the presence of an additional scalar, $\rho$, that mixes with the
Higgs boson. Then, the allowed annihilation channels are now duplicated:
being at a resonance with $h$ and/or $\rho$, or at high masses,
where annihilations into $h$ and/or $\rho$ pairs may open up. The
last case, when $h$ and/or $\rho$ are (a bit) more massive than
the pseudo-scalar is known as forbidden DM. Moreover, the relic abundance
can also be reproduced in the limit of small mixing, via annihilations
into $h$ and/or $\rho$ pairs. Annihilations in $\rho$ pairs when
$m_{\rho}<m_{\theta}$ provide an example of secluded DM.

Our analysis shows that these \textit{minimal models} may potentially
be distinguished among themselves if a positive signal in direct detection
is observed. Measuring the DM mass (say, by a gamma ray line in indirect
detection), for instance, can yield further information regarding
the underlying symmetry. Moreover a positive detection of a new scalar
would yield further information on the mixing, $s_{\alpha}$, and
on the symmetry breaking scale. It should have couplings to SM fermions
like those of the Higgs boson but suppressed by the mixing with respect
to the latter, with possible extra decay modes $\rho\rightarrow\theta\theta$
and/or $\rho\rightarrow hh$ if kinematically allowed.

In addition to this, we have analyzed the possibility of light DM. At the $\rho$ resonance the lowest DM mass could be at the sub-GeV scale where annihilations into muons and hadrons are still possible, whereas in the forbidden/secluded DM cases the mass could be even lower, specially in the forbidden scenario for which indirect detection bounds do not apply.

We also study the case in which two of these symmetry breaking terms
appear simultaneously, to see if the parameter space increases. In
this case, the dark CP symmetry, which for just one symmetry breaking
term (that can always be taken as real) was preserved also after SSB
of the $U(1)$, may be violated by the vacuum. We have derived the
restrictions for all possible pair combinations of the symmetry breaking
terms that appear in the \textit{minimal models} so that the dark
CP is preserved after SSB and, therefore, the pseudo-scalar is stable.
We find that, once these DM stability constraints are taken into account,
the allowed parameter space opens up precisely in the region between
the two \textit{minimal models}. Therefore, we can conclude that adding
more symmetry breaking terms (at least by pairs) fills up the parameter
space between the \textit{minimal models}.

Finally, for small explicit symmetry breaking compared to the scale
of SSB of the $U(1)$, we have obtained an effective low energy Lagrangian
including both the usual Higgs portal as well as a derivative Higgs
portal. This effective Lagrangian turns out to be very useful in order
to analyze DM phenomenology, specifically the relic abundance in the
case of the Higgs boson resonance and direct detection. 

\section*{Acknowledgements}

The authors would like to thank David Cerdeño for showing interest regarding light dark matter which motivated us to further look into it. This work is partially supported by the grant FPA2017-84543-P funded by FEDER/MCIyU-AEI, grant PID2020-113334GB-I00 funded by\\MCIN/AEI/10.13039/501100011033 and grant PROMETEO/2019/087 funded by the ``Generalitat Valenciana". JHG is supported by the ``Generalitat Valenciana" through the GenT Excellence Program (CIDEGENT/2020/020) and grants PID2020-113334GB-I00 and PID2020-113644GB-I00 funded by MCIN/AEI/10.13039/501100011033. LC and CF are also supported by the ``Generalitat
Valenciana" under the ``GRISOLIA" and ``ACIF" fellowship programs,
respectively. 

\appendix

\section{\label{sec:Correspondence-re-complex}Correspondence between real
and complex parametrizations}

The correspondence between the couplings in the real and complex parametrization,
Eqs.~\eqref{eq:2scalar-potential} and~\eqref{eq:complex-scalar-potential}
respectively, is shown in Table~\ref{tab:corr}.
\begin{table}[H]
\begin{centering}
\begin{tabular}{|c|c|}
\hline 
Real & Complex\tabularnewline
\hline 
$\omega_{1}$ & $\Real{\mu_{H1}}$\tabularnewline
\hline 
$\omega_{2}$ & $-\,\Imag{\mu_{H1}}$\tabularnewline
\hline 
$\alpha_{1}$ & $\Real{\lambda_{H2}}+\lambda_{HS}$\tabularnewline
\hline 
$\alpha_{12}$ & $-2\,\Imag{\lambda_{H2}}$\tabularnewline
\hline 
$\alpha_{2}$ & $-\Real{\lambda_{H2}}+\lambda_{HS}$\tabularnewline
\hline 
$\delta_{1}$ & $\Real{\mu^{3}}$\tabularnewline
\hline 
$\delta_{2}$ & $-\Imag{\mu^{3}}$\tabularnewline
\hline 
\end{tabular}~%
\begin{tabular}{|c|c|}
\hline 
Real & Complex\tabularnewline
\hline 
$m_{1}^{2}$ & $m_{S}^{2}+\Real{\mu_{S}^{2}}$\tabularnewline
\hline 
$m_{12}^{2}$ & $-\Imag{\mu_{S}^{2}}$\tabularnewline
\hline 
$m_{2}^{2}$ & $m_{S}^{2}-\Real{\mu_{S}^{2}}$\tabularnewline
\hline 
$\mu_{1}$ & $\Real{\mu_{1}+\mu_{3}}$\tabularnewline
\hline 
$\mu_{12}$ & $\Imag{-\mu_{1}-3\mu_{3}}$\tabularnewline
\hline 
$\mu_{21}$ & $\Real{\mu_{1}-3\mu_{3}}$\tabularnewline
\hline 
$\mu_{2}$ & $\Imag{-\mu_{1}+\mu_{3}}$\tabularnewline
\hline 
\end{tabular}~%
\begin{tabular}{|c|c|}
\hline 
Real & Complex\tabularnewline
\hline 
$\lambda_{1}$ & $\Real{\lambda_{2}+\lambda_{4}}+\lambda_{S}$\tabularnewline
\hline 
$\lambda_{2}$ & $\Real{-\lambda_{2}+\lambda_{4}}+\lambda_{S}$\tabularnewline
\hline 
$\beta_{12}$ & $\Imag{-2\lambda_{2}-4\lambda_{4}}$\tabularnewline
\hline 
$\lambda_{12}$ & $-6\,\Real{\lambda_{4}}+2\lambda_{S}$\tabularnewline
\hline 
$\beta_{21}$ & $\Imag{-2\lambda_{2}+4\lambda_{4}}$\tabularnewline
\hline 
\multicolumn{1}{c}{} & \multicolumn{1}{c}{}\tabularnewline
\multicolumn{1}{c}{} & \multicolumn{1}{c}{}\tabularnewline
\end{tabular}
\par\end{centering}
\caption{\label{tab:corr}Correspondence between the couplings in the real
and complex parametrizations.}
\end{table}

\section{\label{sec:beta-coeff}Relevant interactions for the \textit{minimal
models}}

The couplings which are relevant for DM phenomenology can be parametrized
as
\begin{equation}
\mathcal{-L}\supset\frac{1}{2}\left(\beta_{h\theta\theta}\theta^{2}+\beta_{h\rho\rho}\rho^{2}\right)h+\frac{1}{2}\beta_{\rho\theta\theta}\theta^{2}\rho\,,\label{eq:betai_relevant_lagrangian}
\end{equation}
with the coefficients $\beta_{i}$ summarized in Table~\ref{tab:betai_coeff}
for the \textit{minimal models}. Note that 
\begin{equation}
\beta_{\rho\theta\theta}=\beta_{h\theta\theta}\left(\sa\rightarrow-\ca,\,m_{h}^{2}\rightarrow m_{\rho}^{2}\right)
\end{equation}
\begin{table}[H]
\begin{centering}
\begin{tabular}{|c|c|c|}
\hline 
 & Linear & Quadratic\tabularnewline
\hline 
$\beta_{h\theta\theta}$ & $\frac{s_{\alpha}\left(m_{\theta}^{2}-m_{h}^{2}\right)}{v_{s}}$ & $-\frac{m_{h}^{2}s_{\alpha}}{v_{s}}$\tabularnewline
\hline 
$\beta_{\rho\theta\theta}$ & $-\frac{c_{\alpha}\left(m_{\theta}^{2}-m_{\rho}^{2}\right)}{v_{s}}$ & $\frac{m_{\rho}^{2}c_{\alpha}}{v_{s}}$\tabularnewline
\hline 
$\beta_{h\rho\rho}$ & $c_{\alpha}s_{\alpha}\left(\frac{s_{\alpha}\left(m_{h}^{2}+2m_{\rho}^{2}\right)}{v}-\frac{c_{\alpha}\left(m_{h}^{2}-3m_{\theta}^{2}+2m_{\rho}^{2}\right)}{v_{s}}\right)$ & $-\frac{c_{\alpha}s_{\alpha}\left(m_{h}^{2}+2m_{\rho}^{2}\right)\left(vc_{\alpha}-s_{\alpha}v_{s}\right)}{vv_{s}}$\tabularnewline
\hline 
\end{tabular}\medskip{}
\par\end{centering}
\begin{centering}
\begin{tabular}{|c|c|c|}
\hline 
 & Cubic & Quartic\tabularnewline
\hline 
$\beta_{h\theta\theta}$ & $-\frac{s_{\alpha}\left(m_{h}^{2}+m_{\theta}^{2}\right)}{v_{s}}$ & $-\frac{s_{\alpha}\left(m_{h}^{2}+2m_{\theta}^{2}\right)}{v_{s}}$\tabularnewline
\hline 
$\beta_{\rho\theta\theta}$ & $\frac{c_{\alpha}\left(m_{\rho}^{2}+m_{\theta}^{2}\right)}{v_{s}}$ & $\frac{c_{\alpha}\left(m_{\rho}^{2}+2m_{\theta}^{2}\right)}{v_{s}}$\tabularnewline
\hline 
$\beta_{h\rho\rho}$ & $\frac{1}{3}c_{\alpha}s_{\alpha}\left(\frac{3s_{\alpha}\left(m_{h}^{2}+2m_{\rho}^{2}\right)}{v}-\frac{c_{\alpha}\left(3m_{h}^{2}+m_{\theta}^{2}+6m_{\rho}^{2}\right)}{v_{s}}\right)$ & $-\frac{c_{\alpha}s_{\alpha}\left(m_{h}^{2}+2m_{\rho}^{2}\right)\left(vc_{\alpha}-s_{\alpha}v_{s}\right)}{vv_{s}}$\tabularnewline
\hline 
\end{tabular}
\par\end{centering}
\caption{\label{tab:betai_coeff}Expressions for the $\beta_{i}$ coefficients
particularized for the \textit{minimal models} in terms of the physical
parameters $v,\,\protect\vs,\,m_{h},\,\protect\mdm,\,m_{\rho}$ and
$s_{\alpha}$.}
\end{table}

\section{\label{sec:Hadronization-effects}Hadronic decay modes}

We follow the procedure described by Ref.~\citep{Cline:2013gha}
with the cross section for the $\rho$-mediated\footnote{Note that in our notation $\rho$ is the new scalar, not the $\rho$-meson.} s-channel DM annihilations
into hadronic final states written as
\begin{equation}
\sigma v_{rel}=\frac{4\beta_{\rho\theta\theta}^{2}}{\sqrt{s}}\,|D_{\rho}(s)|^{2}\,\Gamma_{\rho\rightarrow\text{\text{hadrons}}}(\sqrt{s})\,,\label{eq:cline1}
\end{equation}
with
\begin{equation}
|D_{\rho}(s)|^{2}=\frac{1}{\left(s-m_{\rho}^{2}\right)^{2}+m_{\rho}^{2}\Gamma_{\rho,{\rm full}}^{2}\left(m_{\rho}\right)}\,.
\end{equation}
The $\beta_{\rho\theta\theta}$ coefficient is given by the Lagrangian
in Eq.~\eqref{eq:betai_relevant_lagrangian}, the decay width of
the $\rho$ into hadrons, $\Gamma_{\rho\rightarrow\text{\text{hadrons}}}$,
is taken from Fig.~4 in Ref.~\citep{Winkler:2018qyg}, and the full
$\rho$ width, $\Gamma_{\rho,{\rm full}}$, is just the $\Gamma_{\rho\rightarrow\theta\theta}$,
which is the dominant one in the considered parameter space. This decay width is described by
\begin{equation}
\Gamma_{\rho\rightarrow\theta\theta}=\frac{\beta_{\rho\theta\theta}^{2}}{32\pi\,m_{\rho}}\sqrt{1-\frac{4m_{\theta}^{2}}{m_{\rho}^{2}}}\,.
\end{equation}

\section{\label{sec:cases-H1-H2}Direct detection for explicit symmetry breaking
terms with the Higgs field}

Here we also consider the following explicit symmetry breaking terms
with the Higgs field,
\begin{align}
V_{H1} & =\frac{1}{2}\mu_{H1}|H|^{2}S+\mathrm{H.c.}\,,\label{eq:complex-scalar-potential-VH1}\\
V_{H2} & =\frac{1}{2}\lambda_{H2}|H|^{2}S^{2}+\mathrm{H.c.}\,.\label{eq:complex-scalar-potential-Vh2}
\end{align}
We focus on the DD constraints (irrespectively of the relic abundance,
as in the first analysis in Sec.~\ref{subsec:Comparison-of-minimal})
for models including these potentials. We provide the effective DM-nucleon
couplings generated in Table~\ref{tab:effDMnucleoncoupling_extras}.
\begin{table}[H]
\begin{centering}
\begin{tabular}{|c|c|}
\hline 
Model & $\lambda_{SI}\propto-\left(\frac{\beta_{h\theta\theta}\,c_{\alpha}}{m_{h}^{2}}+\frac{\beta_{\rho\theta\theta}\,s_{\alpha}}{m_{\rho}^{2}}\right)$\tabularnewline
\hline 
\hline 
$V_{H1}$ & $\frac{s_{\alpha}c_{\alpha}}{\vs m_{h}^{2}m_{\rho}^{2}}m_{\theta}^{2}(m_{h}^{2}-m_{\rho}^{2})-\frac{2m_{\theta}^{2}}{vm_{h}^{2}m_{\rho}^{2}}(s_{\alpha}^{2}m_{h}^{2}+c_{\alpha}^{2}m_{\rho}^{2})$\tabularnewline
\hline 
$V_{H2}$ & $-\frac{2m_{\theta}^{2}}{vm_{h}^{2}m_{\rho}^{2}}(s_{\alpha}^{2}m_{h}^{2}+c_{\alpha}^{2}m_{\rho}^{2})$\tabularnewline
\hline 
\end{tabular}
\par\end{centering}
\caption{\label{tab:effDMnucleoncoupling_extras}Effective DM-nucleon coupling
that enters in the DD cross section in terms of the physical parameters
$v,\,\protect\vs,\,m_{h},\,\protect\mdm,\,m_{\rho}$ and $s_{\alpha}$.}
\end{table}

\begin{figure}[H]
\begin{centering}
\includegraphics[scale=0.3]{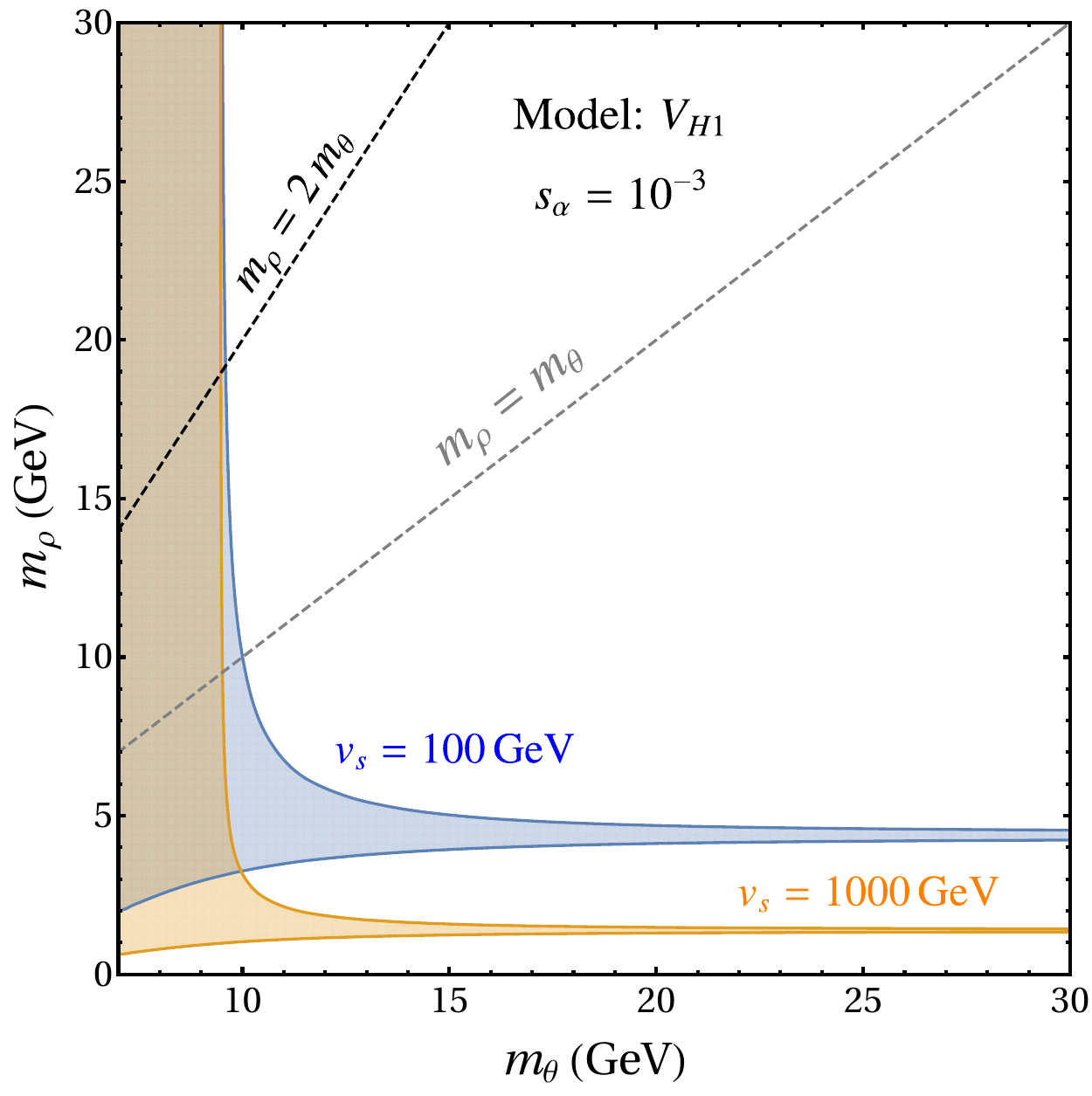}~\includegraphics[scale=0.3]{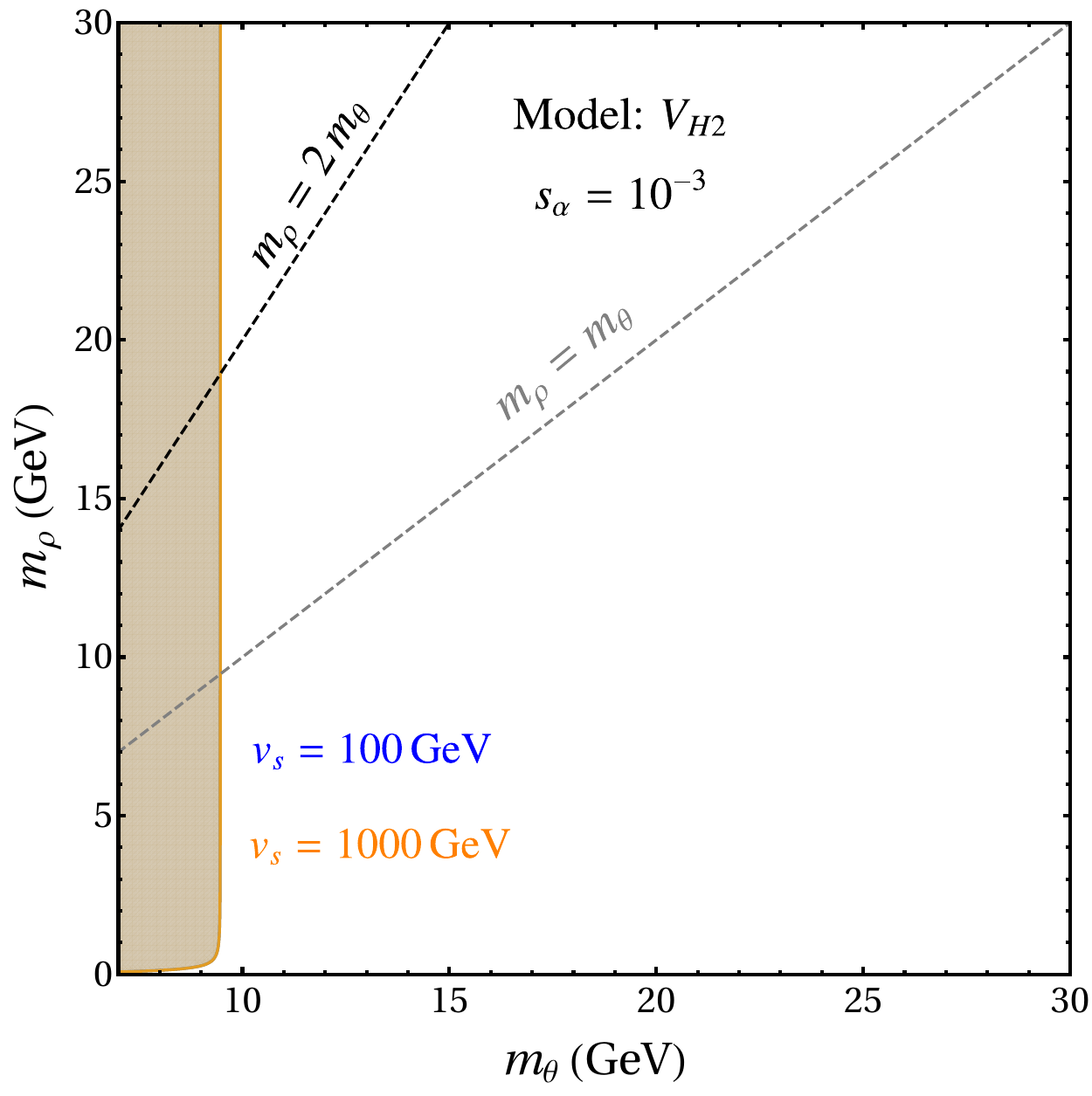}
\par\end{centering}
\caption{\label{fig:DDextra_potential}Parameter space that is allowed by the
XENON1T bound for the models with the symmetry breaking terms in the
potential $V_{H1}$ and $V_{H2}$ in Eqs.~\eqref{eq:complex-scalar-potential-VH1}
and \eqref{eq:complex-scalar-potential-Vh2}. Blue (Orange) colored
region corresponds to $\protect\vs=100\,(1000)\,\protect\gev$. Black
and gray dashed lines represent the resonance condition with the $\rho$
$(m_{\rho}=2m_{\theta})$ and the degenerate case $(m_{\rho}=m_{\theta})$
respectively.}
\end{figure}

Fig.~\ref{fig:DDextra_potential} shows the allowed parameter space
by the XENON1T bound for both models. We plot two values of the VEV
($\vs=10^{2},\,10^{3}\,\gev$ in blue and orange, respectively) and set
$\sa=10^{-3}$, which is the typical value for the mixing in order
to get the correct relic abundance near the resonances (see Fig.~\ref{fig:Z3Z4res}).
We can see that DM masses larger than $10\,\gev$ are excluded by
the experimental constraint from XENON1T.

\setlength{\bibsep}{0pt}

\bibliographystyle{JHEP}
\bibliography{DM_DCP}

\end{document}